\documentclass[preprint,showpacs,amsmath,amssymb,prd]{revtex4}% Physical Review D
\usepackage{epsf}
\usepackage{dcolumn}% Align table columns on decimal point
\usepackage{amsmath}
\usepackage{amssymb}
\usepackage{color}
\usepackage{bm}% bold math
\everymath{\displaystyle}
\begin{document}

\title{Local Lorentz transformations and Thomas effect in general relativity}

\author{Alexander J. Silenko\footnote{Email: alsilenko@mail.ru}}

\affiliation{Research Institute for Nuclear Problems, Belarusian State University, Minsk 220030, Belarus\\
and Bogoliubov Laboratory of Theoretical Physics, Joint Institute for Nuclear Research,
Dubna 141980, Russia}

%\date{\today}

\begin{abstract}
The tetrad method is used for an introduction of local Lorentz frames and a detailed analysis of local Lorentz transformations. A formulation of equations of motion in local Lorentz frames is based on the Pomeransky-Khriplovich gravitoelectromagnetic fields. These fields are calculated in the most important special cases and their local Lorentz transformations are determined.
The local Lorentz transformations and the Pomeransky-Khriplovich gravitoelectromagnetic fields are applied for a rigorous derivation of a general equation for the Thomas effect in Riemannian spacetimes and for a consideration of
Einstein's equivalence principle and the Mathisson force.
\end{abstract}

% 04.20.Cv  Fundamental problems and general formalism
% 04.25.-g  Approximation methods; equations of motion
% 03.65.Sq Semiclassical theories and applications
\maketitle

%\author{Alexander J. Silenko\footnote{Email: alsilenko@mail.ru} \\
%        {\it Research Institute for Nuclear Problems} \\ {\it Belarusian State University, Minsk 220030, Belarus}\\
%{\it Bogoliubov Laboratory of Theoretical Physics} \\ {\it Joint Institute for Nuclear Research,
%Dubna 141980, Russia}}

%%%%%%%%%%%%%%%%%%%%%%%%%%%%%%%%%%%%%%%%%%%%%%%%%
\section{Introduction}
%%%%%%%%%%%%%%%%%%%%%%%%%%%%%%%%%%%%%%%%%%%%%%%%%

Methods of description of gravitational phenomena based on an introduction of tetrads are often used in contemporary gravity. Any tetrad can characterize a local Lorentz frame (LLF) attributed to some observer. The LLFs applied in the classical monograph \cite{MTW} for discussing Einstein's equivalence principle are also very convenient for a description of spin effects including quantum mechanical analysis with the covariant Dirac equation (see, e.g., the reviews \cite{rmp,HehlTwoLectures}).
Since the metric of LLFs is locally Minkowskian, a transition from one LLF (also called a coframe or a tetrad frame) to another one is defined by an appropriate Lorentz transformation. Such Lorentz transformations have been considered in Refs. \cite{Warszawa,OST}.
Basic tetrads satisfy the Schwinger gauge \cite{Schwinger,dirac} (see also \cite{OST}) while other tetrads can also be used. Tetrads which do not satisfy this gauge are carried by observers moving in a described spacetime. For example, a tetrad with $e^{\widehat{0}}_i\neq0,~e^{0}_{\widehat{i}}\neq0$ in a Schwarzschild field is attributed to an observer moving relative to the source. All tetrads are equivalent and the use of any tetrad is possible. Nevertheless, Hamiltonians and equations of motion of a test particle do not coincide even for different tetrads belonging to the Schwinger gauge \cite{Arminjon,Arminjontalk} (see Subsec. \ref{curvilinearcoordinates}).

Great success achieved in description of electromagnetic phenomena stimulated a search for direct analogies between the electrodynamics and gravity.
Indeed, the Newton and Coulomb laws as well as the Coriolis and Lorentz forces seems to be similar. This is the reason why one often applies a conception of gravitoelectromagnetism  based on the introduction of scalar and vector potentials of a gravitoelectromagnetic field. Then one introduces gravitoelectic and gravitomagnetic fields connected each with others by Maxwell-like equations (see Refs. \cite{MashhoonGrEM,Tartaglia,Ryder} and references therein). Since mathematical tools of the electrodynamics and gravity significantly differ, this approach ensures one only an approximate description of gravitational phenomena.

A new big step in formulation of more exact equations of gravitoelectromagnetic fields has been made by Pomeransky and Khriplovich \cite{PK}. They have started from standard equations stating the zero values of covariant de\-ri\-va\-ti\-ves of the four-spin and four-velocity. A following transition to a LLF has allowed them to derive general equations for the tetrad components of the four-spin and four-velocity. The obtained equations are pretty similar to the corresponding equations in electrodynamics, namely, to the Thomas-Bargmann-Michel-Telegdi equation for a Dirac particle and to the equation of motion of a charged particle. This similarity has made it possible to derive general formulas for the gravitoelectromagnetic fields \cite{PK} defined in an anholonomic tetrad frame and describing a \emph{relativistic} particle in an \emph{arbitrarily strong} gravitational field or in a noninertial frame. It is important to mention that this approach is based on the equivalence principle extended on the spin by Kobzarev and Okun \cite{KO} (see also Ref. \cite{T2}).

The most general description of motion of a spinning particle in general relativity (GR) is provided by the Mathisson-Papapetrou (MP) equations \cite{Mathisson,Papapetrou}. These equations predict the violation of the weak equivalence principle for pointlike spinning particles (see Refs. \cite{Plyatsko:1997gs,PlyatskoF,PlyatskoFn} and references therein). Nevertheless, the MP and Pomeransky-Khriplovich (PK) equations agree when one can neglect the mutual influence of
particle and spin motion leading to the aforementioned violation \cite{Warszawa}. This circumstance substantiates the results obtained by Pomeransky and Khriplovich and brings a possibility of a wide application of the approach based on the PK gravitoelectromagnetic fields. However, the original method \cite{PK} used the symmetric gauge which is
inconvenient and may stimulate a wrong interpretation of results obtained (see Refs. \cite{Warszawa,OST}). In particular, the formula for the angular velocity of spin rotation in a rotating frame derived by this method differs from the well-known Gorbatsevich-Mashhoon formula \cite{Mashhoon}. Therefore, we use the Schwinger gauge.

A set of previously obtained results \cite{Warszawa,OST,ostrong,OSTgrav} has shown the applicability of the conception of gravitoelectromagnetism based on the PK gravitoelectromagnetic fields and on the Schwinger gauge. While the PK equations define the general form of the gravitoelectromagnetic fields, the weak-field approximation happens to be rather convenient to obtain simple expressions for the gravitoelectromagnetic fields clarifying their physical meaning. For stationary spacetimes, such expressions have been deduced in Ref. \cite{Warszawa}. In the present work, we derive general formulas for the gravitoelectromagnetic fields in this approximation. The formulas obtained are applicable for a time-dependent metric. We exactly calculate the gravitoelectromagnetic fields in several important special cases. We determine the local Lorentz transformations of these fields. We also use the gravitoelectromagnetic fields for an analysis of fundamental problems of the Mathisson force and Einstein's equivalence
principle.

The very important problem of spin physics is the Thomas precession. The general formula for the Thomas precession in electrodynamics is based on special relativity \cite{Thomas} and is perfectly substantiated \cite{Jackson,DraganThomas,Rindler,Stepanov,PhysScr}. However, manifestations of the Thomas effect in GR are much less clear. In the present work, we fulfill the general description of the Thomas precession in gravity with the use of the local Lorentz transformations and the gravitoelectromagnetic fields.

The paper is organized as follows. The detailed analysis of Lorentz transformations in coframes is carried out in Sec \ref{Lotrfra}.
In Sec.~\ref{Hamiltonian}, we expound the conception based on the PK gravitoelectromagnetic fields and use it for a derivation of the equations of motion in coframes. We calculate the gravitoelectromagnetic fields in the most important special cases in Sec. \ref{examples}. The local Lorentz transformations of these fields are determined in Sec. \ref{Lorentztransformations}. The validity of these transformations is shown in the case of the uniformly accelerated frame. In Sec. \ref{Comparison}, we apply the gravitoelectromagnetic fields for a demonstration of a difference between a behavior of spinning particles in the uniformly accelerated frame and in the Schwarzschild spacetime. The Mathisson force is considered in Sec. \ref{Mtforce}. We rigorously derive the general formula for the Thomas precession in arbitrary Riemannian spacetimes in Sec. \ref{Thomasprecession}.
The results obtained are summarized in Sec.~\ref{final}.

We denote world and spatial indices by Greek and Latin letters
$\alpha,\mu,\nu,\ldots$ $=0,1,2,3,~i,j,k,\ldots=1,2,3$, respectively. Tetrad
indices are denoted by Latin letters from the beginning of the
alphabet, $a,b,c,\ldots = 0,1,2,3$. %%%%%%%%%%%%Separate
Temporal and spatial tetrad indices are
distinguished by hats. The signature is $(+---)$.
%, the Ricci scalar curvature is defined by $R=g^{\mu\nu}R_{\mu\nu}=g^{\mu\nu}R^\alpha_{~\mu\alpha\nu}$, where
%$R^\alpha_{~\mu\beta\nu}=\partial_\beta\Gamma^\alpha_{~\mu\nu}-\ldots$ is the Riemann curvature tensor.
%We use the system of units $\hbar=1,~c=1$ except for some specific expressions.
%
Commas and semicolons before indices denote partial and covariant derivatives, respectively.

\section{Lorentz transformations in coframes}\label{Lotrfra}

Let us consider Lorentz transformations in coframes. Our explanation partially follows Refs. \cite{Warszawa,OST}.

One of the most powerful methods in GR is an introduction of tetrads. They define the LLF characterized by the Minkowski metric $ds^2=\eta_{ab}dx^adx^b, ~ \eta_{ab}={\rm diag}(1, -1, -1, -1)$. The metric tensor of a given spacetime can be split into tetrads $e^a_\mu$ satisfying the relations
\begin{equation}
e^a_\mu e_{a\nu}=g_{\mu\nu},~~~
e_{a\mu} e_b^\mu=\eta_{ab}, ~~~ e^a_\mu e_b^\mu=\delta^a_{b},~~~ e^a_\mu e_a^\nu=\delta^\nu_{\mu}.\label{tetreln}
\end{equation}
As usual, the world and tetrad indices (which all run from 0 to 3) are raised and lowered with
the metric and Minkowski tensors, $g_{\mu\nu}$ and $\eta_{ab}$, respectively.

Any tetrad can be attributed to an observer. Observers carrying different tetrads may move relative to each other. A locality of a Lorentz frame defined by some tetrad is caused by nonzero derivatives of the metric tensor. First derivatives define forces like the Newton force while second derivatives define the spacetime curvature and tidal forces. These forces are felt by the observer and can be detected in the observer's lab. In particular, the velocity of light is equal to $c$ near the observer but the light can be accelerated (due to the Newton-like force) and can undergo a deflection.

In the present work, we apply the approach based on the LLFs for a derivation of the equations of motion of a spinning particle. We use the approximation
disregarding effects conditioned by second derivatives of the metric tensor. In particular, we do not consider the spin-curvature coupling. To study such effects, some other approaches seem to be more convenient. For example, one can describe the spin-curvature coupling with the MP equations (see Ref. \cite{OSTgrav}).

Let us consider two observers and two LLFs in the same area, i.e., in the vicinity of some point $(x_{(0)}^0,x_{(0)}^1,x_{(0)}^2,x_{(0)}^3)$. Since $dx^a=e^a_\mu dx^\mu,~d{x'}^a={e'}^a_\mu dx^\mu$, the connection between coordinates in the two frames is given by \cite{Warszawa}
\begin{equation}
d{x}^a=T^a_b d{x'}^b,
\label{relationsh}
\end{equation} where
\begin{equation}
T^a_b={e}^a_\mu {e'}^\mu_b.
\label{relationT}
\end{equation} We mention that the both tetrads are bound with the same metric tensor
$g_{\mu\nu}=e_{a \mu}e^a_\nu={e'}_{a \mu}{e'}^a_\nu$.

The connection between the four-velocities in the two frames has the form
\begin{equation}
{u}^{a}\equiv \frac{d{x}^a}{d\tau}=T^a_b {u'}^b, \label{relationu}
\end{equation} where $\tau=s/c$ is the proper time.

We should underline that these relations are not valid beyond the local area.

Certainly, the connection between two LLFs is realized by local Lorentz transformations. If the axes in the two frames are parallel and the direction of the velocity $\bm V$ of the primed frame in the unprimed one is arbitrary, this connection is given by
\begin{equation}\begin{array}{c}
d{x'}^{\widehat{0}}=\gamma(dx^{\widehat{0}}-\bm\beta\cdot d\widehat{\bm r}),~~~ d\widehat{\bm r}\,'=d\widehat{\bm r}+\frac{\gamma^2}{\gamma+1}\bm\beta(\bm\beta\cdot d\bm r)-\bm\beta\gamma dx^{\widehat{0}},\\
d{x}^{\widehat{0}}=\gamma(d{x'}^{\widehat{0}}+\bm\beta\cdot d\widehat{\bm r}\,'),~~~ d\widehat{\bm r}=d\widehat{\bm r}\,'+\frac{\gamma^2}{\gamma+1}\bm\beta(\bm\beta\cdot d\widehat{\bm r}\,')+\bm\beta\gamma d{x'}^{\widehat{0}},
~~~\bm\beta=\frac{\bm V}{c}, \end{array}\label{Ltr}
\end{equation}  where $\gamma=(1-\bm\beta^2)^{-1/2}$ is the Lorentz factor. We need to specify that the axes of the two frames remain to be parallel.
Unlike the usual Lorentz transformations, the relative motion of the LLFs may be accelerated ($\dot{\bm V}\neq0$).

The present analysis demonstrates that the quantities $d{x}^a$ and ${u}^{a}$ are four-vectors relative to the local Lorentz transformations. The four-momentum $p_a=-\partial S/(\partial x^a)$ ($S$ is an action) and other four-vectors with tetrad components possess the same property. The tetrad $e_a^\mu$ is a world four-vector when $a$ is fixed and is a four-vector relative to the local Lorentz transformations when $\mu$ is fixed.

However, we should note that the quantity $e_a^\mu A^a$ may not be a covariant four-vector even if $A^a$ is a four-vector relative to the local Lorentz transformations. Examples of such a situation are given in Sec. IVD of Ref. \cite{OSTORSION}.

It is important to consider the case when the unprimed frame is at rest relative the world one. In this case, the unprimed tetrad satisfies the Schwinger gauge
\cite{dirac,Schwinger,Warszawa,OST} ($e^{\widehat{0}}_i=0,~e^{0}_{\widehat{i}}=0$). We can mention that the Schwinger gauge defines an infinite set of tetrads carrying by observers immobile in the world frame. These tetrads are connected by \emph{spatial} transformations. Different tetrads (satisfying the Schwinger gauge) lead to different equations of motion \cite{Arminjon}. The corresponding Hamiltonians also differ \cite{Arminjon,Arminjontalk}. However, appropriate coordinate transformations establish connections between them.

Equation (\ref{relationT}) reduces in the weak-field approximation when gravitational and noninertial fields are weak [$|g_{\mu\nu}-\eta_{\mu\nu}|\ll1~(\mu,\nu=0,1,2,3)$]. Since the quantities $e_{\widehat{0}}^0$ and $e_{\widehat{i}}^i$ are close to 1, the equation $e^{a}_{\mu} e_a^\nu=\delta_{\mu}^\nu=0~(\mu\neq\nu)$ results in $e^{\widehat{\nu}}_{\mu}=-e_{\widehat{\mu}}^{\nu}$. Here hats point out the tetrad indices. The following relations are valid:
\begin{equation}
g_{0i}=e^{\widehat{0}}_i-e_0^{\widehat{i}},~~~ g_{0i}={e'}^{\widehat{0}}_i -
{e'}_0^{\widehat{i}}={e'}_{\widehat{0}}^i -
{e'}^0_{\widehat{i}},~~~
T^{\widehat{0}}_{\widehat{i}}=e^{\widehat{0}}_i +
{e'}^0_{\widehat{i}}, ~~~
T_{\widehat{0}}^{\widehat{i}}=e_0^{\widehat{i}} +
{e'}_{\widehat{0}}^i.
\label{Valid}\end{equation}
Evidently, $T^{\widehat{0}}_{\widehat{i}}-T_{\widehat{0}}^{\widehat{i}}=0$.

Equation (\ref{Ltr}) can be presented in the form
\begin{equation}\begin{array}{c}
d{x'}^{a}=L_b^a d{x}^b, \end{array}\label{Ltrfo}\end{equation} where $L_b^a$ is the Lorentz tensor.
Therefore, \begin{equation}\begin{array}{c}
T^a_b=L_b^a, \end{array}\label{Ltrft}\end{equation}
where \begin{equation}\begin{array}{c}
L_{\widehat{0}}^{\widehat{0}}=\gamma, ~~~ L_{\widehat{i}}^{\widehat{0}}=L_{\widehat{0}}^{\widehat{i}}=-\beta^i\gamma, ~~~
L_{\widehat{i}}^{\widehat{j}}=\delta_{i}^{j}+\frac{\gamma^2}{\gamma+1}\beta^i\beta^j. \end{array}\label{eqdop}\end{equation}

Equations (\ref{relationT}), (\ref{Ltr}), (\ref{Ltrft}), and (\ref{eqdop}) define the dependence of the relative motion of observers on the tetrads carrying by them.

The connection between the two tetrads can also be obtained in an explicit form. Equation (\ref{Ltrfo}) and the definition of tetrads result in
\begin{equation}\begin{array}{c}
{e'}_\mu^{a}=L_b^a e^b_\mu. \end{array}\label{Ltbtt}\end{equation}

The tetrads can be used even for two observers moving in the Minkowski spacetime. If we suppose the first observer to be at rest, the tetrad carried by him has only trivial components. In this case, $dx^{\widehat{\mu}}=dx^{\mu}$ and Eq. (\ref{Ltbtt}) takes the form
\begin{equation}\begin{array}{c}
{e'}^{a}_{\mu}=L_\mu^a,~~~
{e'}^{\widehat{0}}_0=\gamma, ~~~ {e'}^{\widehat{0}}_i={e'}^{\widehat{i}}_0=-\beta^i\gamma,~~~ {e'}^{\widehat{i}}_j=\delta^{i}_j+\frac{\gamma^2}{\gamma+1}\beta^i\beta^j.
\end{array}\label{LtrMink}\end{equation}
The tetrad (\ref{LtrMink}) satisfies the requirements (\ref{tetreln}).

Let us also consider an accelerated frame. The metric tensor is given by
$$g_{\mu\nu}=\left(\left[1+\frac{\bm a\cdot\bm r}{c^2}\right]^2,-1,-1,-1\right),$$
where $\bm a$ is an acceleration. For the observer at rest, the only nontrivial tetrad component is $e^{\widehat{0}}_0=1+\bm a\cdot\bm r/c^2$.
To simplify the analysis, we can suppose that the second observer moves in the LLF of the first observer along the $x^1$ axis with the velocity $V$ and the first observer is at rest in the world frame. In this case, the nontrivial tetrad components are \begin{equation}\begin{array}{c}
{e'}^{\widehat{0}}_0=\gamma{e}^{\widehat{0}}_0,~~~{e'}^{\widehat{1}}_0=-\beta\gamma{e}^{\widehat{0}}_0,~~~{e'}^{\widehat{0}}_1=-\beta\gamma, ~~~{e'}^{\widehat{1}}_1=\gamma.
\end{array}\label{accefra}\end{equation}

%%%%%%%%%%%%%%%%%%%%%%%%%%%%%%%%%%%%%%%%%%%%%%%%%
\section{Equations of motion in coframes}\label{Hamiltonian}
%%%%%%%%%%%%%%%%%%%%%%%%%%%%%%%%%%%%%%%%%%%%%%%%%

In the present work, we explain and develop the conception of gravitoelectromagnetism first proposed by Pomeransky and Khriplovich \cite{PK}. This conception has been advanced in several works \cite{Warszawa,OST,ostrong,OSTgrav,OSTORSION,Dvornikov,Obzor}.

Certainly, there are other conceptions of gravitoelectromagnetism. The conventional conception of gravitoelectromagnetism (see Refs. \cite{MashhoonGrEM,Tartaglia,Ryder}) is based on a four-potential of gravitoelectromagnetic field expressed in terms of components of the metric tensor.
This conception can be applied for a nonrelativistic particle in the weak-field approximation and does not work in the relativistic case. Of course, this is nothing but the simplest way to introduce the gravitoelectromagnetic fields. Some other approaches provide a description of dynamics of spinning particles beyond the nonrelativistic approximation and the weak-field one. We can mention the known papers by Bailey and Israel \cite{BaileyIsrael} and by Yee and Bander \cite{YeeBander}. Important results have been obtained \cite{SteinhoffSchafer} with the canonical formulation of GR based on the Arnowitt-Deser-Misner parametrization. In this connection, Ref. \cite{BarausseRacineBuonanno} can also be noticed. The results obtained allow one to draw a parallel between electromagnetism and gravity. In particular, some approaches connecting this parallel have been proposed in Refs. \cite{Jantzen,Costa,Schafer}. Established relations between electromagnetism and gravity are of interest. Nevertheless, only the gravitoelectromagnetic fields introduced by Pomeransky and Khriplovich allow one to reveal a deep analogy between spinning particles in electromagnetic and gravitational/inertial fields.

We can note that a canonical method has also been applied in Refs. \cite{OST,ostrong,OSTgrav,OSTORSION,PRD,PRD2007} for a derivation of quantum mechanical equations of motion. In Ref. \cite{OSTgrav}, this method has also been used to obtain the corresponding classical equations. In the present work, we basically follow the original method by Pomeransky and Khriplovich \cite{PK}.
The distinctive feature of the PK fields is their introduction in the coframes.

The equation of motion of a pointlike spinless particle in GR is given by
\begin{equation}
\begin{array}{c}
Du^{\mu}= 0.
\end{array} \label{eqPKu} \end{equation}
This equation defines the particle motion on a geodesic line which is perturbed by the Mathisson force for spinning particles and by tidal forces for extended ones. As a rule, the Mathisson and tidal forces are relatively small and can be neglected in the present study. The motion of spinning particles with allowance for a particle deflection from the geodesic line is defined by the Mathisson-Papapetrou equations \cite{Mathisson,Papapetrou}.

The orthogonality condition interconnects the spin four-vector $a^\mu$ with either the four-velocity or the four-momentum. The three-component spin $\bm\zeta$ is defined in the particle rest frame, when $a^{\mu}=(0,\bm\zeta)$. %There are two forms of the orthogonality condition.
The Mathisson-Pirani \cite{Mathisson,Pirani} condition connects the spin with the four-velocity,
\begin{equation}
\begin{array}{c}
u^{\mu}a_\mu= 0,
\end{array} \label{eqMatPi} \end{equation}
while the Tulczyjev condition \cite{T} joins the spin with the four-momentum,
\begin{equation}
\begin{array}{c}
p^{\mu}a_\mu= 0.
\end{array} \label{eqMatTu} \end{equation}
A choice of specific condition does not influence the next derivations. However, we can note that there exist situations when
the results following from the MP equations with the supplementary condition (\ref{eqMatTu}) are not satisfactory from the physical point of view \cite{Plyatsko2015}.

When Eq. (\ref{eqPKu}) is satisfied,
\begin{equation}
\begin{array}{c}
Du^{a}= D(u^{\mu}e^a_{\mu})= u^{\mu}e^a_{\mu;\nu}dx^\nu.
\end{array} \label{eqPK1} \end{equation}
Therefore,
\begin{equation}
\begin{array}{c}
\frac{Du^{a}}{d\tau}= e^a_{\mu;\nu}u^{\mu}u^\nu =
e_b^\mu e^a_{\mu;\nu}e_c^\nu u^{b}u^c =\Gamma^a_{~bc}u^bu^c.
\end{array} \label{eqPK2} \end{equation}
Here $\Gamma_{abc}=-\Gamma_{bac}= e_b^\mu e_c^\nu e_{a\mu;\nu}$ are the Lorentz connection coefficients (Ricci rotation coefficients). They can also be presented in the form
\begin{equation}\begin{array}{c}
\Gamma_{abc}=\frac 12\left(\lambda_{abc}+\lambda_{bca}-\lambda_{cab}\right),~~~
\lambda_{abc}=-\lambda_{acb}=e_b^\mu e_c^\nu (e_{a\mu,\nu}-e_{a\nu,\mu}).
\end{array}\label{eqin7}\end{equation}

Since $u^{a}= u^{\mu}e^a_{\mu}$ is a world scalar, $Du^{a}=d u^{a}$ and Eq. (\ref{eqPK2}) takes the form \cite{PK}
\begin{equation}
\begin{array}{c}
\frac{du^{a}}{d\tau}=\Gamma^{a}_{~bc} u^b u^c.
\end{array} \label{eqPK3} \end{equation}

Since $D(u^{\mu}a_\mu)=0$, Eqs. (\ref{eqPKu}) and (\ref{eqMatPi}) result in
\begin{equation}
\begin{array}{c}
Da^{\mu}= 0.
\end{array} \label{eqPKn} \end{equation}

Equations (\ref{eqPKu}) and (\ref{eqPKn}) can be considered as a mathematical formulation of the equivalence principle for the particle and spin.

The same derivation as above leads to the equation of spin motion in LLFs. Since
\begin{equation}
\begin{array}{c}
\frac{Da^{a}}{d\tau}= e^a_{\mu;\nu}u^{\mu}u^\nu =
e_b^\mu e^a_{\mu;\nu}e_c^\nu u^{b}u^c =\Gamma^a_{~bc}a^bu^c,
\end{array} \label{eqPKspn} \end{equation}
we finally obtain
\begin{equation}
\begin{array}{c}
\frac{da^{a}}{d\tau}= \Gamma^a_{~bc}a^bu^c.
\end{array} \label{eqPKspin} \end{equation}

Equations (\ref{eqPK3}) and (\ref{eqPKspin}) have been first obtained in Ref. \cite{PK}. We can note that the use of LLFs for a description of spin dynamics is quite natural. The three-component spin which evolution is commonly described is defined in the particle rest frame. This frame can be attributed to some observer and the local Lorentz transformation of the spin (pseudo)vector coincides with its transformation in special relativity (see, e.g., Ref. \cite{LL4}):
\begin{equation}
a^a=(a^{\widehat{0}},~{\widehat{\bm a}}),~~~{\widehat{\bm a}}=\bm{\zeta}+\frac{\gamma^2{\bm\beta}({\bm\beta}\cdot\bm{\zeta})}{\gamma+1},~~~a^{\widehat{0}}={\bm\beta}\cdot{\widehat{\bm a}}=\gamma{\bm\beta}\cdot\bm{\zeta}.
\label{spin} \end{equation}
In this case,
\begin{equation}
u^a=(u^{\widehat{0}},\widehat{\bm u})=(\gamma,\gamma{\bm\beta}).
\label{momentum} \end{equation} Let us remember that $\bm V=\bm\beta c$ is the velocity of relative motion of the two LLFs. Therefore, the quantity $\bm{\zeta}$ defines here the three-component spin in the instantaneously accompanying frame. The spin motion in the nonrotating instantaneously accompanying frame and in the particle rest frame differs due to the Thomas effect \cite{Thomas}. This difference is defined by \cite{Thomas,Jackson}
\begin{equation}   \begin{array}{c}
\left(\frac{\partial\bm\zeta}{\partial t}\right)_{nonrot}=\left(\frac{\partial\bm\zeta}{\partial t}\right)_{rest\,frame}+\bm\omega_T\times\bm\zeta,
\end{array} \label{Thomprecession} \end{equation}
where $\omega_T$ is the angular velocity of the Thomas precession. In electrodynamics and special relativity,
\begin{equation}\begin{array} {c}
\bm\omega_T=-\frac{\gamma}{\gamma+1}\left(\bm \beta\times\frac{d\bm \beta}{d\tau}\right).
\end{array}\label{Thompre}\end{equation}

Equations (\ref{eqPK3}) and (\ref{eqPKspin}) are similar to the corresponding equations of motion in electrodynamics:
\begin{equation}
\begin{array}{c}
\frac{du_{\mu}}{d\tau}=\frac{e}{m}F_{\mu\nu} u^\nu,
\end{array} \label{Loreq} \end{equation}
\begin{equation}
\begin{array}{c}
\frac{ds_{\mu}}{d\tau}=\frac{e}{m}F_{\mu\nu} s^\nu.
\end{array} \label{eqThBMT} \end{equation}
Equation (\ref{eqThBMT}) describes the spin motion of a particle with the \emph{Dirac magnetic moment} ($g=2$).

The electromagnetic field tensor can be expressed in terms of the electric and magnetic fields, $F_{\mu\nu}=(\bm E,\bm B)$. A similarity between $(e/m)F_{\mu\nu}$ and $\Gamma_{abc}u^c$ has allowed Pomeransky and Khriplovich to introduce the gravitoelectric and gravitomagnetic fields, $c\Gamma_{abc}u^c=(\bm{\mathcal{E}},\bm{\mathcal{B}})$. Explicitly \cite{PK}
\begin{equation}
\begin{array}{c} \mathcal{E}_{\widehat{i}}=c\Gamma_{0ic}u^{c},~~~
\mathcal{B}_{\widehat{i}}=-\frac{c}{2}e_{ikl}\Gamma_{klc}u^{c}.
\end{array} \label{expl}
\end{equation} For the gravitoelectromagnetic fields $\bm{\mathcal{E}}$ and $\bm{\mathcal{B}}$, we do not make a difference between upper and lower indices.

As a result of comparison of the foregoing equations of motion in gravitational/inertial and electromagnetic fields, Pomeransky and Khriplovich \cite{PK} have obtained the following equation of motion for the three-component spin:
\begin{equation}
\begin{array}{c} \frac{d\bm\zeta}{d\tau}=\bm{\Omega}'\times \bm\zeta, ~~~ \bm{\Omega}'=-\bm{\mathcal{B}}+
\frac{\widehat{\bm u}\times\bm{\mathcal{E}}}
{u^{\widehat{0}}+1}.
\end{array} \label{omgemtau}
\end{equation}

The time dilation is given by
\begin{equation}
dt=u^0d\tau, ~~~ d\widehat{t}=u^{\widehat{0}}d\tau.
\label{timedil}
\end{equation}
As a result,
\begin{equation}
\begin{array}{c} \frac{d\bm\zeta}{dt}=\bm{\Omega}\times \bm\zeta, ~~~ \bm{\Omega}=\frac{1}{u^0}\left(-\bm{\mathcal{B}}+
\frac{\widehat{\bm u}\times\bm{\mathcal{E}}}
{u^{\widehat{0}}+1}\right).
\end{array} \label{omgem}
\end{equation}
The validity of Eqs. (\ref{omgemtau}) and (\ref{omgem}) has been substantiated in Refs. \cite{Warszawa,Dvornikov,OST}. It has been demonstrated \cite{Warszawa,OST} that the LLF which is at rest relative to the world frame satisfies the Schwinger gauge. Other gauges define the equations of motion in frames moving relative to the world frame. This important property was not taken into account in Refs. \cite{PK,Obzor,Dvornikov}.

The equations of particle motion can be presented as follow \cite{PK}:
\begin{equation}
\begin{array}{c} \frac{d\widehat{\bm{u}}}{d\tau}=u^{\widehat{0}}\bm{\mathcal{E}}+
\widehat{{\bm
u}}\times\bm{\mathcal{B}},  ~~~ \frac{d{u}^{\widehat{0}}}{d\tau}=
\bm{\mathcal{E}}\cdot\widehat{\bm u}.
\end{array} \label{force}
\end{equation}

We can underline that the equation of motion (\ref{eqPK3}) is essentially nonlinear and Eq. (\ref{eqPKspin}) contains the four-velocity. These properties differ the equations of motion in GR from the corresponding equations in electrodynamics [Eqs. (\ref{Loreq}) and (\ref{eqThBMT})]. Therefore, the gravitoelectromagnetic fields depend on the four-velocity and they are effective fields. In particular, one cannot use these fields to construct a free gravitoelectromagnetic field.

Since the particle velocity in the coframe is equal to $\widehat{\bm v}\equiv d\widehat{\bm r}/d\widehat{t}=c\widehat{\bm u}\left/\sqrt{1+{\widehat{\bm u}}^2}\right.$ and $d/d\widehat{t}=(1/{u}^{\widehat{0}})(d/d\tau)$, the corresponding acceleration is given by
\begin{eqnarray}
\widehat{\bm w}\equiv\frac{d\widehat{\bm v}}{d\widehat{t}}=\frac {c}{{u}^{\widehat{0}}}\left[\frac{d\widehat{\bm u}}{d\widehat{t}}-\frac{\widehat{\bm u}}{\left({u}^{\widehat{0}}\right)^2}\left(\widehat{\bm u}\cdot \frac{d\widehat{\bm u}}{d\widehat{t}}\right)\right]
=\frac {c}{{u}^{\widehat{0}}}\left[\bm{\mathcal{E}} %\right.\nonumber\\ \left.
+\widehat{\bm{\beta}} \times\bm{\mathcal{B}}-\widehat{\bm{\beta}} \left(\bm{\mathcal{E}}\cdot\widehat{\bm{\beta}}\right)\right],\label{RotFFor}
\end{eqnarray} where $\widehat{\bm{\beta}}=\widehat{\bm v}/c$.

We should take into account that Eqs. (\ref{eqPKspin}),  (\ref{omgem}), and (\ref{force}), (\ref{RotFFor}) are not equally useful. Equations (\ref{eqPKspin}) and (\ref{omgem}) are perfect for a description of the spin motion. It is often admissible to disregard an inhomogeneity of the gravitoelectromagnetic fields and their dependence on the four-velocity. One can also take into account the aforementioned inhomogeneity and the evolution of the four-velocity in the LLF. As a contrary, Eqs. (\ref{force}) and (\ref{RotFFor}) do not define measurable dynamics of the four-velocity. First of all, one cannot ignore the nonlinearity %of Eq. (\ref{omgem})
caused by the dependence of the gravitoelectromagnetic fields on the four-velocity. Second, an observer needs a description of dynamics of the four-velocity in the world frame. In particular, the conventional force which governs the particle motion is defined in this frame and has the form $f^i=mc(du^i/dt)$.
%\ref{LL}.
An example of the description of the particle motion in the world frame will be presented in Sec. \ref{Comparison}.

Thus, the introduction of the gravitoelectromagnetic fields \emph{defined in LLFs} simplifies the description of dynamics of spinning particles in GR. Equation (\ref{expl}) shows that the important specific feature of the gravitoelectromagnetic fields is their dependence on the particle four-velocity. However, the results presented do not define some other fundamental properties of gravitoelectromagnetic fields. It is very important to find a transformation law of these fields. This problem will be solved in Sec. \ref{Lorentztransformations}.

%%%%%%%%%%%%%%%%%%%%%%%%%%%%%%%%%%%%%%%%%%%%%%%%%
\section{Gravitoelectromagnetic fields in some important special cases}\label{examples}
%%%%%%%%%%%%%%%%%%%%%%%%%%%%%%%%%%%%%%%%%%%%%%%%%

It is instructive to calculate the gravitoelectromagnetic fields in the most important special cases. Certainly, we will determine these fields in the LLFs which are at rest relative to the corresponding world frames. We will consider, among others, several examples of a nonstationary (time-dependent) metric, the importance of which for analyzing fundamental problems of GR has been recently affirmed in Ref. \cite{Arminjon}.

The general form of the line element of an
arbitrary gravitational field can be given by \cite{ostrong,hergt}
\begin{equation}\label{LT}
ds^2 = V^2c^2dt^2 - \delta_{\widehat{i}\widehat{j}}W^{\widehat
i}{}_k W^{\widehat j}{}_l\,(dx^k - K^kcdt)\,(dx^l - K^lcdt).
\end{equation}

An analysis of the most important metrics can be simplified with the use of the isotropic (more exactly, Cartesian-like isotropic) coordinates. In this case, the line element takes the form (see Ref. \cite{OST})
\begin{equation}\label{LTOST}
ds^2 = V^2c^2dt^2 - W^2\,\delta_{ij}\,(dx^i - K^icdt)\,(dx^j - K^jcdt).
\end{equation}

\subsection{General noninertial frame}

The accelerated and rotating noninertial frame presents the general case of inertial fields in a flat spacetime. The acceleration ${\bm a}$ and the
angular velocity of rotation $\bm\omega$ of an observer are
independent of the spatial coordinates but may depend
arbitrarily on time. The {\it exact} metric of the general noninertial frame has the form (\ref{LTOST}), where
\begin{equation}
V = 1 + {\frac {{\bm a(t)}\cdot{\bm r}}{c^2}},~~~ W = 1,~~~
\bm K =-{\frac 1c}\,(\bm\omega(t)\times\bm r).\label{VWni}
\end{equation}

Explicitly, the metric is given by \cite{HN}
\begin{equation}\label{gennoni}
ds^2 =\left[\left(1+\frac{\bm a(t)\cdot\bm r}{c^2}\right)^2-\frac{[\bm\omega(t)\times\bm r]^2}{c^2}\right]c^2dt^2 - 2[\bm\omega(t)\times\bm r]\cdot d\bm rdt - \delta_{ij}dx^idx^j.
\end{equation}

The nonzero Lorentz connection coefficients have the same forms for time-dependent and time-independent inertial fields,
\begin{equation}\begin{array}{c}
\Gamma_{\widehat{i}\widehat{j}\widehat{0}}=-\frac{ce_{ijk}\omega^k(t)}{c^2+\bm a(t)\cdot\bm r}, ~~~ \Gamma_{\widehat{0}\widehat{i}\widehat{0}}=-\Gamma_{\widehat{i}\widehat{0}\widehat{0}}=-\frac{a^i(t)}{c^2+\bm a(t)\cdot\bm r}.
\end{array}\label{Rirok}\end{equation}

The gravitoelectromagnetic fields are given by
\begin{equation}\begin{array}{c}
\bm{\mathcal{E}}=-\frac{c\bm a(t)}{c^2+\bm a(t)\cdot\bm r}u^{\widehat{0}}, ~~~ \bm{\mathcal{B}}=\frac{c^2\bm\omega(t)}{c^2+\bm a(t)\cdot\bm r}u^{\widehat{0}}.
\end{array}\label{gemfi}\end{equation}
There is only the gravitoelectric field in the uniformly accelerated frame,
\begin{equation}\begin{array}{c}
\bm{\mathcal{E}}=-\frac{c\bm a(t)}{c^2+\bm a(t)\cdot\bm r}u^{\widehat{0}}, ~~~ \bm{\mathcal{B}}=0.
\end{array}\label{gefua}\end{equation}
In the rotating frame, there is only the gravitomagnetic field:
\begin{equation}\begin{array}{c}
\bm{\mathcal{E}}=0, ~~~ \bm{\mathcal{B}}=\bm\omega(t)u^{\widehat{0}}.
\end{array}\label{gmrfm}\end{equation} It is important that the gravitoelectric field does not depend on $\widehat{\bm u}$.
All equations presented in this subsection are \emph{exact}.

%Evidently, Eqs. (\ref{gemfi}), (\ref{gefua}), and (\ref{gmat rfm}) agree with the nonrelativistic and weak-field approximation ensured by the conventional approach %[Eq. (\ref{gemfipt})].
The use of Eqs. (\ref{omgem}) and (\ref{force}) allows one to reproduce known formulas for the particle motion and the spin rotation in the general noninertial frame.

\subsection{Cylindrical coordinate system}\label{curvilinearcoordinates}

When the cylindrical coordinate system is used, the spacetime is flat. However, the metric tensor is nontrivial and has the form $g_{\mu\nu}={\rm diag}(1, -1, -\rho^2, -1)$. The simplest tetrad satisfying the Schwinger gauge has the only nontrivial component $e_\phi^{\widehat{\phi}}\equiv e_2^{\widehat{2}}=\rho$. As a result, the nonzero Lorentz connection coefficients are
\begin{equation}\Gamma_{\widehat{2}\widehat{1}\widehat{2}}=-\Gamma_{\widehat{1}\widehat{2}\widehat{2}}=\frac1\rho.
\label{Riccinz}\end{equation}

The gravitoelectromagnetic fields are given by \cite{SurfInv}
\begin{equation}
\begin{array}{c} \bm{\mathcal{E}}=0,~~~
\mathcal{B}_{\widehat{\rho}}\equiv \mathcal{B}_{\widehat{1}}=0, ~~~ \mathcal{B}_{\widehat{\phi}}\equiv \mathcal{B}_{\widehat{2}}=0,~~~\mathcal{B}_{\widehat{z}}\equiv \mathcal{B}_{\widehat{3}}=\frac{u^{\widehat{2}}}{\rho}=u^\phi\equiv u^2.
\end{array} \label{expcylf}
\end{equation}

Equations (\ref{force}) and (\ref{expcylf}) show that the force determined by the gravitomagnetic field acting in the
cylindrical coordinate system is an analogue of
the Lorentz force. Its appearance is a consequence of
the fact that, if the azimuthal angle of the particle
changes by $d\widehat{\phi}$, the horizontal axes of the cylindrical
and Cartesian systems of coordinates rotate by the
same angle with respect to each other. Thus, the cylindrical coordinate system rotates with an instantaneous angular velocity $-d\widehat{\phi}/dt=-v_{\widehat{\phi}}/\rho$ with respect to the Cartesian one \cite{SurfInv}.

The nonzero gravitomagnetic field leads to forces acting on particles and torques rotating spins. However, these forces and torques are fictitious. Their appearance is caused by the fact that particle trajectories have different shapes in the Cartesian and cylindrical coordinate systems. The  aforementioned forces and torques are not felt by an observer. In contrast, the observer feels the acceleration force and the centrifugal one acting in the general noninertial frame.

The result can differ for another tetrad even if it also satisfies the Schwinger gauge. Let us consider an observer using the Cartesian coordinates ($x^{\widehat{0}}=x^0,~x^{\widehat{1}}=\rho\cos{\phi},~x^{\widehat{2}}=\rho\sin{\phi},~x^{\widehat{3}}=z$). In this case, the nontrivial tetrad components are given by
$${e}^{\widehat{1}}_1=\cos{\phi},~~~{e}^{\widehat{1}}_2=-\rho\sin{\phi},~~~{e}^{\widehat{2}}_1=\sin{\phi}, ~~~{e}^{\widehat{2}}_2=\rho\cos{\phi}.$$
It is easy to obtain that the gravitoelectromagnetic fields are equal to zero ($\bm{\mathcal{E}}=0, ~\bm{\mathcal{B}}=0$).
This result is natural for the Cartesian coordinate system in the Minkowski spacetime.

This comparison of the two tetrads satisfying the Schwinger gauge elucidates the statement made in Ref. \cite{Arminjon}. A tetrad field in the Schwinger gauge is not unique and different tetrads may lead to different equations of motion. The corresponding Hamiltonians do not coincide either \cite{Arminjontalk,SurfInv}.
%%%%%%% to summary %%%%%%%%%%%%%%%%%%%%%%%%%%%%%%%%%%%%%%%%%%%%%%%%%%%%%%%%%%%%%%%%%%%%%%%%%%%%%%%%%%%%%%%%%%%%%%%%%%%%%%%%%%%%%%%%%%%%%%%%%%%%%%%%%%%%%%%%%%%%%%%%%%%%%%
However, the forces and torques caused by a difference of the tetrads are fictitious and are not felt by an observer.

\subsection{Gravitoelectromagnetic fields in the weak-field approximation}

The weak-field approximation can often be used. To find gravitoelectromagnetic fields in this approximation, we use the Schwinger gauge and suppose that the tetrad components $e_\mu^{\widehat{\mu}}~(\mu=\widehat{\mu},~\mu=0,1,2,3)$ are close to unit.
Under these conditions, some important general relations can be obtained,
\begin{equation}\begin{array}{c}
e_{\widehat{\mu}}^\nu+e_{\mu}^{\widehat{\nu}}=\delta_{\mu}^{\nu}, ~~~ g_{\mu\nu}=e_{\widehat{\mu}\nu}+e_{\widehat{\nu}\mu},\\
\Gamma_{abc}=\frac12\left(e_{\widehat{a}b,c}-e_{\widehat{b}a,c}+ g_{bc,a}-g_{ac,b}\right).
\end{array}\label{genlrel}\end{equation}

As a result, nonzero Lorentz connection coefficients are given by
\begin{equation}\begin{array}{c}
\Gamma_{\widehat{0}\widehat{i}\widehat{0}}=-\frac12g_{00,i}, ~~~ \Gamma_{\widehat{0}\widehat{i}\widehat{j}}=-\frac12\left(g_{0i,j}+g_{0j,i}-g_{ij,0}\right), ~~~ \Gamma_{\widehat{i}\widehat{j}\widehat{0}}=\frac12\left(g_{0j,i}-g_{0i,j}\right),\\
\Gamma_{\widehat{i}\widehat{j}\widehat{k}}=\frac12\left(g_{jk,i}-g_{ik,j}\right).
\end{array}\label{Lotzcncff}\end{equation}

The gravitoelectromagnetic fields are equal to
\begin{equation}
\begin{array}{c} \mathcal{E}_{\widehat{i}}=-\frac{c}{2}\left[g_{00,i}u^{\widehat{0}}+
\left(g_{0i,j}+
g_{0j,i}-g_{ij,0}\right)u^{\widehat{j}}\right],\\
\mathcal{B}_{\widehat{i}}=\frac{c}{4}e_{ijk}\left[(g_{0j,k}-g_{0k,j})
u^{\widehat{0}}+\left(g_{jl,k}-g_{kl,j}\right)u^{\widehat{l}}\right].
\end{array} \label{explg}
\end{equation}

In Ref. \cite{Warszawa}, Eqs. (\ref{Lotzcncff}) and (\ref{explg}) have been obtained for a stationary metric. Only the gravitoelectric field explicitly depends on the time derivative.

We can conclude that the equations of motion in coframes (\ref{omgem}) and (\ref{force}) become very simple in the weak-field approximation. In this case, they contain only first derivatives of the metric tensor and can be easily derived. The equations obtained are relativistic.
%%%%%%% to summary %%%%%%%%%%%%%%%%%%%%%%%%%%%%%%%%%%%%%%%%%%%%%%%%%%%%%%%%%%%%%%%%%%%%%%%%%%%%%%%%%%%%%%%%%%%%%%%%%%%%%%%%%%%%%%%%%%%%%%%%%%%%%%%%%%%%%%%%%%%%%%%%%%%%%%
%%%They show a great similarity between gravity and electromagnetism.

\subsection{Lense-Thirring metric}\label{LenThir}

Lense and Thirring
have discovered in 1918 that rotating bodies ``drag'' the spacetime around
themselves (frame dragging \cite{LT}). In other words, they have demonstrated
the similarity between rotating frames and spacetimes created by rotating
bodies.

The Lense-Thirring (LT) metric \cite{LT} defines a gravitational field of a rotating source in the weak-field approximation. It can be obtained from the Kerr metric when the distance from the source is much large than the gravitational radius. The static part of the LT metric characterizes the Schwarzschild field of a distant source. It is convenient to transform the LT metric to the isotropic coordinates \cite{OST},
\begin{equation}\begin{array}{c}
V =1 - {\frac {GM}{c^2r}}, ~~~
W=1 + {\frac {GM} {c^2r}},~~~
\bm K=\frac {\bm\omega\times\bm r}{c},\\
\bm\omega =\frac{2G}{c^2r^3}\bm J= \left(0,~ 0,~
\frac{2GMa}{c\,r^3}\right).\label{omega}
\end{array}\end{equation}
Here $\bm J=Mca\bm e_z$ is the total angular momentum of the source and $M$ is its mass.

For this metric, the gravitoelectromagnetic fields read
\begin{equation}\begin{array}{c}
\bm{\mathcal{E}}=-\frac{Gm}{cr^3}\bm ru^{\widehat{0}}+\frac{3G}{c^2r^5}\left[{\bm r}
({\bm J}\cdot({\bm r}\times\widehat{\bm u})) -({\bm r}\times{\bm J})(\widehat{\bm u}\cdot{\bm r})\right], \\ \bm{\mathcal{B}}=-\frac{Gm}{cr^3}\bm r\times\widehat{\bm u}-\frac{G}{c^2r^3}\left [\frac{3(\bm r\cdot\bm J) \bm r}{r^2}-\bm J\right]u^{\widehat{0}}.
\end{array}\label{gemLenT}\end{equation}
Because (see Ref. \cite{OST})
$$\begin{array}{c}
({\bm r}\times\widehat{\bm u})({\bm J}\cdot({\bm r}\times\widehat{\bm u})) +(\widehat{\bm u}\times({\bm r}\times{\bm J}))(\widehat{\bm u}\cdot{\bm r})=
2({\bm r}\times\widehat{\bm u})({\bm J}\cdot({\bm r}\times\widehat{\bm u})) \\ +
(\widehat{\bm u}\times({\bm r}\times\widehat{\bm u}))({\bm r}\cdot{\bm J})+r^2(\widehat{\bm u}\times(\widehat{\bm u}\times{\bm J})),\end{array}$$
equations (\ref{omgem}) and (\ref{gemLenT}) lead to the following equivalent equations of spin motion \cite{OST}:
\begin{equation}\begin{array}{c}\label{OmegaGK}
{\bm \Omega} =\frac {1}{u^0}\Biggl\{\frac {GM}{cr^3}\cdot\frac {2u^{\widehat{0}}+1}{u^{\widehat{0}}+1}{\bm r}\times\widehat{\bm u}+
\frac {G}{c^2r^5}\left[3{\bm r}({\bm r}
\cdot{\bm J})-r^2\bm J\right]u^{\widehat{0}}\\-\frac {3G}{c^2r^5}\cdot\frac {1}{u^{\widehat{0}}+1}\left[({\bm r}\times\widehat{\bm u})\left({\bm J}\cdot({\bm r}\times\widehat{\bm u})\right)+
({\bm r}\cdot\widehat{\bm u})\left(\widehat{\bm u}\times({\bm r}\times{\bm J})\right)\right]\Biggr\},
\end{array}\end{equation}
\begin{equation}\begin{array}{c}\label{OmegaGL}
{\bm \Omega} =\frac {1}{u^0}\Biggl\{\frac {GM}{cr^3}\cdot\frac {2u^{\widehat{0}}+1}{u^{\widehat{0}}+1}{\bm r}\times\widehat{\bm u}+
\frac {G}{c^2r^5}\left[3{\bm r}({\bm r}
\cdot{\bm J})-r^2\bm J\right]u^{\widehat{0}}\\-\frac {3G}{c^2r^5}\cdot\frac {1}{u^{\widehat{0}}+1}\left[2({\bm r}\times\widehat{\bm u})\left({\bm J}\cdot({\bm r}\times\widehat{\bm u})\right)+
(\widehat{\bm u}\times({\bm r}\times\widehat{\bm u}))({\bm r}\cdot{\bm J})+r^2(\widehat{\bm u}\times(\widehat{\bm u}\times{\bm J}))\right]\Biggr\}.
\end{array}\end{equation}
These equations describe the geodetic precession and the LT one for the \emph{relativistic} particle.

Equations (\ref{OmegaGK}) and (\ref{OmegaGL}) agree with the approximate formula by
%%%%%%%Lense and Thirring \cite{LT}.
However, one of important preferences of the relativistic approach is the discovery of additional dependence of $\bm{\mathcal{E}}$ and $\bm{\mathcal{B}}$ on the nondiagonal and diagonal components of the metric tensor, respectively. The corresponding contributions to the angular velocity of the spin precession do not follow from the nonrelativistic approximation.

\subsection{Static gravitational fields in isotropic coordinates}

The static metric in the isotropic coordinates is defined by Eq. (\ref{LTOST}) at condition that $\bm K=0$ and has the form
\begin{equation}\label{LTisotr}
ds^2 = V^2c^2dt^2 - W^2(d\bm r\cdot d\bm r).
\end{equation}

The respective gravitoelectromagnetic fields are given by
\begin{equation}\begin{array}{c}
\bm{\mathcal{E}}=-cu^{\widehat{0}}\nabla V, ~~~ \bm{\mathcal{B}}=c\,\nabla W\times\widehat{\bm u}.
\end{array}\label{gemaisot}\end{equation}

An appearance of nonzero $\bm{\mathcal{B}}$ depending on $W$ is a
new property as compared with the nonrelativistic approximation.

The most important examples of static fields are the Schwarzschild, de Sitter, and anti-de Sitter spacetimes. We use the weak-field approximation.

For the Schwarzschild metric in the isotropic coordinates, $V$ and $W$ are given by Eq. (\ref{omega}). In the weak-field approximation, the gravitoelectromagnetic fields take the form
\begin{equation}\begin{array}{c}
\bm{\mathcal{E}}=-\frac{GM\bm r}{cr^3}u^{\widehat{0}}=\frac{\bm gu^{\widehat{0}}}{c}, ~~~ \bm{\mathcal{B}}=-\frac{GM}{cr^3}\bm r\times\widehat{\bm u}=\frac{\bm g\times\widehat{\bm u}}{c},
\end{array}\label{gemfSch}\end{equation} where $\bm g$ is the Newtonian acceleration. We can state the significant difference between the gravitoelectromagnetic fields in the uniformly accelerated frame and in the Schwarzschild spacetime. The gravitomagnetic field in the Schwarzschild spacetime, contrary to the uniformly accelerated frame, is nonzero.

The four-dimensional de Sitter metric can be presented in the form
\begin{equation}\begin{array}{c}
ds^2=\left(1-\frac{r^2}{\alpha^2}\right)c^2dt^2-\left(1-\frac{r^2}{\alpha^2}\right)^{-1}dr^2-r^2(d\theta^2+\sin^2{\theta}d\phi^2). \end{array}\label{deSitme}
\end{equation}
There is a cosmological horizon at $r=\alpha$.

The de Sitter spacetime is an Einstein manifold since the Ricci tensor is proportional to the metric:
$$R_{\mu\nu}= \frac{4}{\alpha^2}g_{\mu\nu}.$$
This means that the de Sitter spacetime is a vacuum solution of the Einstein equation with the cosmological constant $\Lambda=3/\alpha^2$ and the scalar curvature
$R=4\Lambda=12/\alpha^2$.

The anti-de Sitter metric can be obtained with the
substitution $\alpha\rightarrow ik$.

The well-known coordinate transformation may
reduce de Sitter and anti-de Sitter metrics to isotropic forms. For the de Sitter metric, this transformation is given by
\begin{eqnarray}
r=\frac{\rho}{1+\frac{\rho^2}{4\alpha^2}}.
\label{trdS}
\end{eqnarray}
The metric takes the form
\begin{eqnarray}
%ds^2=\left(1-\frac{\rho^2}{4\alpha^2}\right)^2\left(1+\frac{\rho^2}{4\alpha^2}\right)^{-2}c^2dt^2
%-\left(1+\frac{\rho^2}{4\alpha^2}\right)^{-2}
%\left[d\rho^2+\rho^2(d\theta^2+\sin^2{\theta}d\phi^2)\right].
ds^2=k_-^2k_+^{-2}c^2dt^2
-k_+^{-2}
\left[d\rho^2+\rho^2(d\theta^2+\sin^2{\theta}d\phi^2)\right], ~~~ k_\pm=1\pm\frac{\rho^2}{4\alpha^2}.
\label{isodS}
\end{eqnarray}
As a result, the isotropic Cartesian coordinates can be used.

The gravitoelectromagnetic fields read
\begin{equation}\begin{array}{c}
\bm{\mathcal{E}}=\frac{c}{\alpha^2}\bm\rho u^{\widehat{0}}, ~~~ \bm{\mathcal{B}}=-\frac{c}{\alpha^2}\bm\rho\times\widehat{\bm u}.
\end{array}\label{gemdeSf}\end{equation}

For the anti-de Sitter metric, the gravitoelectromagnetic fields can be obtained with the substitution $\alpha\rightarrow ik$ and are equal to
\begin{equation}\begin{array}{c}
\bm{\mathcal{E}}=-\frac{c}{k^2}\bm\rho u^{\widehat{0}}, ~~~ \bm{\mathcal{B}}=\frac{c}{k^2}\bm\rho\times\widehat{\bm u}.
\end{array}\label{gemAdSf}\end{equation}

An existence of the \emph{gravitomagnetic} field in static spacetimes is very important for an %comparison of
evolution of momentum and spin. This field is not weak, at least for relativistic particles.

%%%%%%%%%%%%%%%%%%%%%%%%%%%%%%%%%%%%%%%%%%%%%%%%%
\section{Local Lorentz transformations of gravitoelectromagnetic fields}\label{Lorentztransformations}
%%%%%%%%%%%%%%%%%%%%%%%%%%%%%%%%%%%%%%%%%%%%%%%%%

The properties of the local Lorentz transformations allow us to determine the general dependence of the gravitoelectromagnetic fields from the choice of a tetrad. The results obtained in Sec. \ref{Lotrfra} and the explicit expression of $\Gamma_{abc}u^c$ in terms of tetrads (\ref{eqin7}) lead to the conclusion that this quantity is an antisymmetric tensor relative to the local Lorentz transformations. Therefore, the gravitoelectric and gravitomagnetic fields, $\bm{\mathcal{E}}$ and $\bm{\mathcal{B}}$, transform like the electric and magnetic ones,
\begin{equation}\begin{array} {c}
\bm{\mathcal{E}}'=\gamma\left[\bm{\mathcal{E}}-\frac{\gamma}{\gamma+1}\bm\beta(\bm\beta\cdot\bm{\mathcal{E}})
+\bm \beta\times\bm{\mathcal{B}}\right],\\
\bm{\mathcal{B}}'=\gamma\left[\bm{\mathcal{B}}-\frac{\gamma}{\gamma+1}\bm\beta(\bm\beta\cdot\bm{\mathcal{B}})
-\bm \beta\times\bm{\mathcal{E}}\right].
\end{array} \label{meffinal} \end{equation}
This property also establishes a great similarity between electromagnetism and gravity.

When the local Lorentz transformations (\ref{meffinal}) are used, one needs to take into account the dependence of the gravitoelectromagnetic fields on the four-velocity. To express $\bm{\mathcal{E}}'$ and $\bm{\mathcal{B}}'$ in terms of the four-velocity in the primed frame, one needs to transform the components of $u^c$ entering into Eq. (\ref{expl}) as follows:
\begin{equation}\begin{array}{c}
{u}^{\widehat{0}}=\gamma({u'\,}^{\widehat{0}}+\bm\beta\cdot \widehat{\bm u}\,'),~~~ \widehat{\bm u}=\widehat{\bm u}\,'+\frac{\gamma^2}{\gamma+1}\bm\beta(\bm\beta\cdot \widehat{\bm u}\,')+\bm\beta\gamma {u'\,}^{\widehat{0}}. \end{array}\label{LLTf}\end{equation}

Let us consider, as an example, the local Lorentz transformations of the gravitoelectromagnetic fields in the uniformly accelerated frame. Let us suppose that the acceleration is constant and has nonzero projections onto the axes $\bm e^1$ and $\bm e^2$ ($\bm a=a^{(1)}\bm e^1+a^{(2)}\bm e^2$). Let us suppose that the second observer moves in the LLF of the first observer along the $x^1$ axis with the velocity $V$ and the first observer is at rest in the world frame. In this case, the nontrivial tetrad components are given by Eq. (\ref{accefra}). The nontrivial components of the inverse tetrad are equal to
\begin{equation}\begin{array}{c}
{e'}_{\widehat{0}}^0=
\frac{\gamma c^2}{c^2+\bm a\cdot\bm r},~~~{e'}_{\widehat{1}}^0=\frac{\beta\gamma c^2}{c^2+\bm a\cdot\bm r},~~~{e'}_{\widehat{0}}^1=\beta\gamma, ~~~{e'}_{\widehat{1}}^1=\gamma
\end{array}\label{acceinv}\end{equation}
and the nonzero Lorentz connection coefficients ($\Gamma_{abc}=-\Gamma_{bac}$) read
\begin{equation}\begin{array}{c}
\Gamma_{{\widehat{1}}{\widehat{0}}{\widehat{0}}}=\frac{\gamma a^{(1)}}{c^2+\bm a\cdot\bm r},~~~
\Gamma_{{\widehat{2}}{\widehat{0}}{\widehat{0}}}=\frac{\gamma^2 a^{(2)}}{c^2+\bm a\cdot\bm r},~~~
\Gamma_{{\widehat{1}}{\widehat{0}}{\widehat{1}}}=\frac{\beta\gamma a^{(1)}}{c^2+\bm a\cdot\bm r},\\
\Gamma_{{\widehat{0}}{\widehat{2}}{\widehat{1}}}=-\frac{\beta\gamma^2 a^{(2)}}{c^2+\bm a\cdot\bm r},~~~
\Gamma_{{\widehat{2}}{\widehat{1}}{\widehat{0}}}=\frac{\beta\gamma^2 a^{(2)}}{c^2+\bm a\cdot\bm r},~~~
\Gamma_{{\widehat{1}}{\widehat{2}}{\widehat{1}}}=-\frac{\beta^2\gamma^2 a^{(2)}}{c^2+\bm a\cdot\bm r}.
\end{array}\label{eqn12}\end{equation}

The use of Eqs. (\ref{expl}) and (\ref{LLTf}) results in
\begin{equation}\begin{array}{c}
{\mathcal{E}}'_1=-\frac{a^{(1)}c}{c^2+\bm a\cdot\bm r}u^{\widehat{0}}, ~~~ {\mathcal{E}}'_2=-\frac{\gamma a^{(2)}c}{c^2+\bm a\cdot\bm r}u^{\widehat{0}},~~~
{\mathcal{B}}'_3=\frac{\beta\gamma a^{(2)}c}{c^2+\bm a\cdot\bm r}u^{\widehat{0}}.
\end{array}\label{gemfiua}\end{equation}
Since the same result can be easily obtained with Eqs. (\ref{gefua}) and (\ref{meffinal}), this derivation confirms the validity of the general equation (\ref{meffinal}).

The results presented in this section explicitly show the possibility of the local Lorentz transformations of the gravitoelectromagnetic fields and the equivalence of all tetrads. However, the tetrads belonging the Schwinger gauge are much more convenient.

\section{Connection between equations of motion in the uniformly accelerated frame and in the Schwarzschild field and its relation to Einstein's equivalence principle}\label{Comparison}

In this section, we use the results presented for a comparison of motion of a spinning particle in the uniformly accelerated frame and in the Schwarzschild field. This problem is directly related to Einstein's equivalence principle which is a cornerstone of GR. In this section, we do not make a difference between $x$ with upper and lower indices.

\subsection{Previously obtained results}

Equation (\ref{gefua}) and (\ref{gemfSch}) shows the difference between the gravitoelectromagnetic fields in the uniformly accelerated frame and in the Schwarzschild metric in the isotropic coordinates. The corresponding angular velocities of the spin precession in the nonrelativistic limit ($v\ll c$) are equal to \cite{Lee,MashhoonObukhov}
\begin{equation}\begin{array}{c}
\bm\Omega^{(a)}=-\frac{\bm a\times \widehat{\bm v}}{2c^2}, ~~~ \bm\Omega^{(i)}=\frac{3\bm g\times \widehat{\bm v}}{2c^2}.
\end{array}\label{equnLee}\end{equation}
Owing to the difference between the two angular velocities on the condition that $\bm a=-\bm g$, it has been claimed in Refs. \cite{VarjuRyder,Lee} that Einstein's equivalence principle is violated. This claim has been based on a noncoincidence of the quantum mechanical Hamiltonians for a nonrelativistic Dirac particle in the uniformly accelerated frame \cite{HN} and in the Schwarzschild metric in the isotropic coordinates. The latter Hamiltonian has been first derived in Ref. \cite{DH}. In Ref. \cite{VarjuRyder}, the nonrelativistic quantum mechanical Hamiltonians has been obtained for the Schwarzschild metric in the \emph{Cartesian} coordinates but its derivation contains an error. This error has been corrected in Ref. \cite{MashhoonObukhov}.

The Schwarzschild metric in the Cartesian coordinates has the form (\ref{LT}) where
\begin{equation}\begin{array}{c}
 V=1-\Phi,~~~ W^{\widehat
i}{}_k=\delta^i_k+\frac{\Phi x^ix_k}{r^2}, ~~~\bm K=0, ~~~ \Phi={\frac {GM} {c^2r}}.
\end{array}\label{wip}\end{equation}

The appropriate Schwinger tetrad is given by \cite{OSTgrav}
\begin{equation}\begin{array}{c}
 e^{\widehat
0}{}_\mu=V\delta^0_\mu, \quad e_\mu^{\widehat
i} =
W^{\widehat i}{}_k\delta^k_\mu.
\end{array}\label{wipp}\end{equation}

The corrected Hamiltonian for the Schwarzschild metric in the Cartesian coordinates \cite{MashhoonObukhov} leads to the angular velocity of the spin precession coinciding with that defined by the Donoghue-Holstein Hamiltonian \cite{DH},
\begin{equation}\begin{array}{c}
\bm\Omega^{(C)}=\frac{3\bm g\times\widehat{\bm v}}{2c^2}.
\end{array}\label{eqnnLee}\end{equation} In Ref. \cite{MashhoonObukhov}, the relativistic expression for the angular velocity of the spin precession has also been obtained.

Therefore, the angular velocity of the spin precession of the nonrelativistic Dirac particle is three times bigger in the Schwarzschild field than in the uniformly accelerated frame. It has been shown in Ref. \cite{MashhoonObukhov} that the part of the effect in the Schwarzschild spacetime caused by the temporal component of the metric
is equal to the total effect in the accelerated frame and the additional effect in the Schwarzschild spacetime is due to the spatial components of the metric. The classical equations for the spin precession in the Schwarzschild spacetime \cite{Schiff,MashhoonObukhov} and in the uniformly accelerated frame \cite{MashhoonObukhov,Nicolaevici} fully agree with the corresponding quantum mechanical ones. In the general case, the perfect agreement between classical and quantum mechanical equations of motion has been proven in Refs. \cite{OSTgrav,OSTORSION}.

In Ref. \cite{PRD}, the known results has been generalized to a relativistic Dirac particle. The relativistic Foldy-Wouthuysen transformation has been performed and quantum mechanical and semiclassical equations of motion for the momentum and spin have been derived. It has been shown that all equations of motion are different for particles in the Schwarzschild spacetime and in the uniformly accelerated frame. The semiclassical equations of motion in
the Schwarzschild field are given by \cite{PRD}
\begin{eqnarray}
\frac{d\bm p}{dt}=
\frac{2\gamma^2-1}{\gamma}m\bm g,~~~
\frac{d\bm\zeta}{dt}=-\frac{2\gamma
+1}{c^2(\gamma +1)}\left(\bm g\times\bm
v\right)\times\bm\zeta, \label{spinm}\end{eqnarray} where $\bm p=-\{p_i\}$ is the generalized momentum. The corresponding equations  of motion for the accelerated frame read \cite{PRD}
\begin{equation}
\frac{d\bm p}{dt}= -\gamma m\bm a, ~~~ \frac{d\bm\zeta}{dt}=
\frac{\gamma}{c^2(\gamma +1)}\left(\bm a\times\bm
v\right)\times\bm\zeta.
\label{acfraeq32}\end{equation} The weak-field approximation and the isotropic coordinates are used in Eqs. (\ref{spinm}) and (\ref{acfraeq32}).
When $\bm a=-\bm g$, these equations significantly differ.

Equation (\ref{spinm}) and (\ref{acfraeq32}) demonstrate a difference between the particle motion in the Schwarzschild spacetime and in the uniformly accelerated frame.
For example, the light deflection in the Schwarzschild field in the isotropic coordinates ($\bm g=-\bm a$) defined by Eq. (\ref{spinm}) seems to be twice as much as in the uniformly
accelerated frame. This problem will be considered below in details.
An origin of the aforementioned effects is the nonzero spatial part of the Schwarzschild metric. Equations (\ref{gefua}) and (\ref{gemfSch}) show that the gravitoelectric field in the uniformly accelerated frame is the same as in the Schwarzschild spacetime. However, the spatial part of the Schwarzschild metric generates the gravitomagnetic field $\bm{\mathcal{B}}=\bm g\times\widehat{\bm u}/c$ which is absent in the uniformly accelerated frame.

Nevertheless, we do not share the statement about the violation of Einstein's equivalence principle presented in Refs. \cite{VarjuRyder,Lee}. It is incorrect to suppose that the equivalence principle as formulated by Einstein and successors states the complete equivalence of static gravitational fields and uniformly accelerated frames. In Einstein's papers \cite{Einstein},
the equivalence principle has been formulated only relative to constant \emph{uniform} gravitational fields. The Schwarzschild field (as well as other real gravitational fields) is \emph{nonuniform}. Since the equivalence principle is one of fundamental principles of GR, we will consider the problem of the importance of a field inhomogeneity in detail.

We should mention that the presence of tidal and Mathisson forces always differs the static
gravitational field from the uniformly accelerated frame. The tidal forces are
proportional to derivatives of the Newtonian acceleration, i.e.
to %curvature
\emph{second} derivatives of the metric tensor. The Mathisson force, which defines the spin-curvature coupling and is also proportional to
second derivatives of the metric tensor, will be considered in the next section.

\subsection{Comparison of equations of motion in the Schwarzschild field in the Cartesian and isotropic coordinates}

To demonstrate an influence of a spatial inhomogeneity of the Schwarzschild field on the particle motion, we can compare the equations of motion in the Cartesian and isotropic coordinates.
To establish a difference in these equations of motion, it is sufficient to consider a field of a distant source and to use the weak-field approximation. In this approximation, the metric tensors of the Schwarzschild field can be given by \cite{LL2}
\begin{equation}
g^{(C)}_{00}=1-\frac{r_g}{r},\quad g^{(C)}_{0i}=0,\quad g^{(C)}_{ij}=-\left(\delta_{ij}-\frac{r_gx_ix_j}{r^3}\right)
\label{mtCar}\end{equation}
and
\begin{equation}
g^{(i)}_{00}=1-\frac{r_g}{r},\quad g^{(i)}_{0i}=0,\quad g^{(i)}_{ij}=-\left(1-\frac{r_g}{r}\right)\delta_{ij}
\label{mtiso}\end{equation}
in the Cartesian and isotropic coordinates, respectively. Here $r_g=2GM/c^2$ is the gravitational radius.

The use of the general equation (\ref{explg}) for the metric (\ref{mtCar}) results in the following expressions for the gravitoelectromagnetic fields:
\begin{equation}\begin{array}{c}
\bm{\mathcal{E}}^{(C)}=-\frac{cr_g\bm r}{2r^3}u^{\widehat{0}}=\frac{\bm gu^{\widehat{0}}}{c}, ~~~ \bm{\mathcal{B}}^{(C)}=-\frac{cr_g}{2r^3}\bm r\times\widehat{\bm u}=\frac{\bm g\times\widehat{\bm u}}{c}.
\end{array}\label{gemCSch}\end{equation}

A comparison of Eqs. (\ref{gemfSch}) and (\ref{gemCSch}) shows that the gravitoelectromagnetic fields %for the Schwarzschild metric
in the Cartesian coordinates are the same as in the isotropic ones. Therefore, the equations of the spin motion in the Schwarzschild field have the same form in the Cartesian and isotropic coordinates. However, the opposite situation takes place for the particle motion. While the metrics (\ref{mtCar}) and (\ref{mtiso}) characterize the gravitational field of the same source, they belong to different kinds of the spatial inhomogeneity. As a result, the equations of the particle motion in the Cartesian and isotropic coordinates significantly differ.

Let us use the conventional equations of the particle motion
\begin{equation}
\begin{array}{c}
\frac{du_{\mu}}{ds}-\frac12g_{\nu\lambda,\mu}u^\nu u^\lambda=0
\end{array} \label{Lorcova} \end{equation} and
\begin{equation}
\frac{du^{\mu}}{ds}+\left\{^{\,\mu}_{\nu\lambda}\right\}u^\nu u^\lambda=0,
 \label{LorChrc} \end{equation}
where
\begin{equation}
\begin{array}{c}
\left\{^{\,\mu}_{\nu\lambda}\right\}=\frac12g^{\mu\rho}\left(g_{\rho\nu,\lambda}+g_{\rho\lambda,\nu}-g_{\nu\lambda,\rho}\right)
\end{array} \label{Chrcgen} \end{equation} are the Christoffel symbols. The equations of the particle motion in the Cartesian coordinates take the form
\begin{equation}
\begin{array}{c}
\frac{du_{i}}{ds}=\frac{(u^0)^2r_g}{2r^3}\left\{x^i\left[1+\frac{3(\bm\beta\cdot\bm r)^2}{r^2}\right]-2\beta^i(\bm\beta\cdot\bm r)\right\}\\=-\frac{(u^0)^2}{c^2}\left\{g^i\left[1+\frac{3(\bm\beta\cdot\bm r)^2}{r^2}\right]-2\beta^i(\bm\beta\cdot\bm g)\right\},\qquad \frac{du_{0}}{ds}=0,
\end{array} \label{LorcCar} \end{equation}
\begin{equation}
\begin{array}{c}
\frac{du^{i}}{ds}=-\frac{(u^0)^2r_g}{2r^3}x^i\left[1+2\bm\beta^2-\frac{3(\bm\beta\cdot\bm r)^2}{r^2}\right]=\frac{(u^0)^2}{c^2}g^i\left[1+2\bm\beta^2-\frac{3(\bm\beta\cdot\bm r)^2}{r^2}\right],\\ \frac{du^{0}}{ds}=-\frac{(u^0)^2r_g(\bm\beta\cdot\bm r)}{r^3}=2\frac{(u^0)^2}{c^2}(\bm\beta\cdot\bm g).
\end{array} \label{LorcCap} \end{equation} We do not make a difference between the upper and lower indices for the Newtonian acceleration $\bm g$.

The corresponding equations in the isotropic coordinates read
\begin{equation}
\begin{array}{c}
\frac{du_{i}}{ds}=\frac{(u^0)^2r_g}{2r^3}x^i\left(1+\bm\beta^2\right)=-\frac{(u^0)^2}{c^2}g^i\left(1+\bm\beta^2\right),\qquad \frac{du_{0}}{ds}=0,
\end{array} \label{Lorciso} \end{equation}
\begin{equation}
\begin{array}{c}
\frac{du^{i}}{ds}=-\frac{(u^0)^2r_g}{2r^3}\left[x^i(1+\bm\beta^2)-2\beta^i(\bm\beta\cdot\bm r)\right]=\frac{(u^0)^2}{c^2}\left[g^i(1+\bm\beta^2)-2\beta^i(\bm\beta\cdot\bm g)\right],\\ \frac{du^{0}}{ds}=-\frac{(u^0)^2r_g(\bm\beta\cdot\bm r)}{r^3}=2\frac{(u^0)^2}{c^2}(\bm\beta\cdot\bm g).
\end{array} \label{Lorcisp} \end{equation}

These equations can be compared with the related equations for the uniformly accelerated frame,
\begin{equation}
\begin{array}{c}
\frac{du_{i}}{ds}=\frac{(u^0)^2a^i}{c^2},\qquad \frac{du_{0}}{ds}=0,
\end{array} \label{Loraiso} \end{equation}
\begin{equation}
\begin{array}{c}
\frac{du^{i}}{ds}=-\frac{(u^0)^2a^i}{c^2},\qquad \frac{du^{0}}{ds}=-2\frac{(u^0)^2}{c^2}(\bm\beta\cdot\bm a).
\end{array} \label{Loraisp} \end{equation}

The comparison of Eqs. (\ref{LorcCar}) -- (\ref{Loraisp}) shows that the terms different for the Schwarzschild field and the uniformly accelerated frame ($\bm g=-\bm a$) are of the same order of magnitude as the corresponding terms different for the Schwarzschild field in the Cartesian and isotropic coordinates. We can conclude that  the spatial inhomogeneity significantly influences the form of the equations of motion. Therefore, the results presented do not give a %sufficient
reason for the assertion about a violation of Einstein's equivalence principle.

We should mention that the forces conditioned by the mismatched terms in Eqs. (\ref{LorcCar}) -- (\ref{Loraisp}) are proportional to \emph{first} derivatives of the spatial components of the metric tensor.
\section{Mathisson force}\label{Mtforce}
%%%%%%%%%%%%%%%%%%%%%%%%%%%%%%%%%%%%%%%%%%%%%%%%%

One of Mathisson’s great achievements was the discovery of an additional force acting on a spinning particle in a curved spacetime.
The Mathisson force is similar to the Stern-Gerlach one in electrodynamics. The gravitoelectromagnetism allows us to explain the main difference between the two forces as a result of a specific dependence of the gravitoelectromagnetic fields on the particle four-velocity. An analysis of the Mathisson force can be fulfilled in the general case (see, e.g., Refs. \cite{Plyatsko:1997gs,PlyatskoF,PlyatskoFn,OSTgrav,Plyatsko2015}).
Nevertheless, a derivation of a simple expression in the weak-field approximation seems to be rather important. In particular, this expression shows that a  similarity between the gravity and electromagnetism exists even for effects depending on curvature.

To take into account the influence of the spin on the particle motion, we may use the Hamiltonian method and may add the Hamiltonian of a spinless particle in a gravitational field \cite{Cogn} by the term $\bm\zeta\cdot\bm\Omega$,
\begin{equation}
\mathcal{H}=\mathcal{H}_0+\bm\zeta\cdot\bm\Omega.
\label{Hamiltn}\end{equation}
The possibility of this addition has been mentioned in Ref. \cite{PK}. In Refs. \cite{ostrong,OSTgrav}, this addition has been rigorously substantiated not only in framework of classical gravity but also for Dirac particles.

It follows from Eqs. (\ref{omgem}) and (\ref{Hamiltn}) that the additional force acting on a spinning particle is given by
\begin{equation}
\bm f_M=-\nabla(\mathcal{H}-\mathcal{H}_0)=\nabla\left(\frac{1}{u^0}\bm\zeta\cdot\left[\bm{\mathcal{B}}-
\frac{\widehat{\bm u}\times\bm{\mathcal{E}}}
{u^{\widehat{0}}+1}\right]\right).
\label{Matforc}\end{equation}
In electrodynamics, the Stern-Gerlach force \emph{acting on a Dirac particle} ($g=2$) has a %very
similar form,
\begin{equation}
\bm f_{SG}=\frac{e}{mc}\nabla\left(\frac{1}{u^0}\bm\zeta\cdot\left[\bm{B}-
\frac{\bm u\times\bm{E}}
{u^0+1}\right]\right).
\label{SGforce}\end{equation} Equation (\ref{SGforce}) is obtained in the classical limit.
The most important difference between Eqs. (\ref{Matforc}) and (\ref{SGforce}) consists of the dependence of the gravitoelectromagnetic fields on the four-velocity.

Since the Stern-Gerlach and Mathisson forces are proportional to gradients of scalars, they define the additions only to the electric and gravitoelectric forces, respectively. This property has been proved in Ref. \cite{OSTgrav} for arbitrarily strong gravitational fields.
As a result, $\nabla\times\bm f_{SG}=0, ~\nabla\times\bm f_M=0$.

The Mathisson force violates the weak equivalence principle \cite{Plyatsko:1997gs,PlyatskoF,PlyatskoFn} because particles with different spin directions move on different trajectories. Since the gravitoelectromagnetic fields are proportional to first derivatives of the metric, the Mathisson force is proportional to second derivatives of the metric, i.e., to the curvature (see Ref. \cite{OSTgrav}).

The resulting force acting on a spinning particle in a LLF is given by
\begin{equation}
\begin{array}{c} \bm F=\frac{mcu^{\widehat{0}}}{u^0}\left(\bm{\mathcal{E}}+
\frac{\widehat{{\bm
u}}\times\bm{\mathcal{B}}}{u^{\widehat{0}}}\right)+\bm f_M.
\end{array} \label{fullforce}
\end{equation}

In Ref. \cite{OSTgrav}, the Mathisson force has been obtained in an explicit form. Evidently, Eq. (\ref{fullforce}) presents the next-order approximation as compared with the PK equations.

%%%%%%%%%%%%%%%%%%%%%%%%%%%%%%%%%%%%%%%%%%%%%%%%%
\section{Thomas precession in general relativity}\label{Thomasprecession}
%%%%%%%%%%%%%%%%%%%%%%%%%%%%%%%%%%%%%%%%%%%%%%%%%

Amazingly, the use of the local Lorentz transformations and the PK fields allow us to derive the formula for the Thomas precession in inertial and gravitational fields. For this purpose, we apply the method developed in electrodynamics and presented in Refs. \cite{Jackson,PhysScr}. Equation (\ref{Thomprecession}) is in fact the definition of the Thomas effect. We calculate spin dynamics and separate contributions from the local Lorentz transformations and from the Thomas effect to the angular velocity of the spin precession.

To determine the contribution from the local Lorentz transformations, we need to compare the spin motion in the two LLFs connected with the chosen tetrad and with the %nonrotating
observer instantaneously accompanying the test particle. At a given moment of time, the velocity of the test particle in zero. In the general case, the test particle can be accelerated in this frame. It is more convenient to present Eq. (\ref{omgemtau}) in the form
\begin{equation}
\begin{array}{c} \frac{d\bm\zeta}{d\widehat{t}}=\widetilde{\bm{\Omega}}\times \bm\zeta, ~~~ \widetilde{\bm{\Omega}}=\frac{1}{u^{\widehat{0}}}\left(-\bm{\mathcal{B}}+
\frac{\widehat{\bm u}\times\bm{\mathcal{E}}}
{u^{\widehat{0}}+1}\right).
\end{array} \label{omgemtt}
\end{equation}

%Let us distinguish the instantaneously accompanying frame by primes. It
In the instantaneously accompanying frame, the angular velocity of the spin motion is defined only by the gravitomagnetic field,
\begin{equation}
\widetilde{\bm{\Omega}}^{(0)}=-\bm{\mathcal{B}}^{(0)}.
\label{omgemOm}
\end{equation}
%Evidently, in this case
%\begin{equation}
%{u}^{\widehat{\,0}}=\gamma, ~~~ \widehat{\bm u}=\bm\beta\gamma,~~~{u'}^{\widehat{\,0}}=1, ~~~ \widehat{\bm u}'=0. \label{connn}
%\end{equation}

The connection between the angular velocities of the spin motion in the LLF which relates to the chosen tetrad and in the instantaneously accompanying frame is defined by the time
dilation. As follows from Eqs. (\ref{momentum}), (\ref{timedil}), (\ref{meffinal}), and (\ref{omgemOm}), this connection is given by
\begin{equation}\begin{array}{c}
\widetilde{\bm{\Omega}}_L=\frac{\widetilde{\bm{\Omega}}^{(0)}}{{u}^{\widehat{0}}}=-\frac{\bm{\mathcal{B}}^{(0)}}{u^{\widehat{0}}}=-\bm{\mathcal{B}}+\frac{1}{u^{\widehat{0}}(u^{\widehat{0}}+1)}\widehat{\bm u}(\widehat{\bm u}\cdot\bm{\mathcal{B}})+
\frac{\widehat{\bm u}\times\bm{\mathcal{E}}}
{u^{\widehat{0}}},\\
\bm{\Omega}_L=\frac{u^{\widehat{0}}}{{u}^{0}}\left[-\bm{\mathcal{B}}+\frac{1}{u^{\widehat{0}}(u^{\widehat{0}}+1)}\widehat{\bm u}(\widehat{\bm u}\cdot\bm{\mathcal{B}})+
\frac{\widehat{\bm u}\times\bm{\mathcal{E}}}
{u^{\widehat{0}}}\right].
\end{array} \label{omgemOu}
\end{equation}

Equation (\ref{omgemOu}) would be sufficient for a description of the spin precession if the three-component spin were defined in the instantaneously accompanying frame. However, it is defined in the particle rest frame. As a result, we need to take into account the Thomas precession. For this purpose, we follow Refs. \cite{Jackson,PhysScr}. It is convenient to denote
$${\cal F}_{ab}=c\Gamma_{abc}u^c=(\bm{\mathcal{E}},\bm{\mathcal{B}}).$$
With the use of Eq. (\ref{eqPK3}), Eq. (\ref{eqPKspin}) can be presented in the form
\begin{equation}
\frac{da^a}{d\tau}={\cal F}^{ab}a_b-u^a {\cal F}^{bc}u_{b}a_c-
u^a\frac{du^b}{d\tau}a_b.
\label{BMT1} \end{equation}

The next derivations can be made similarly to Refs. \cite{Jackson,PhysScr}. We use the denotation
\begin{equation}
\Phi^a={\cal F}^{ab}a_b-u^a {\cal F}^{bc}u_{b}a_c.
\label{Jacksong} \end{equation}

Evidently, $\Phi^a=(\Phi^{\widehat{0}},\bm{\widehat{\Phi}})$ is a four-vector relative to the local Lorentz transformations.
Since $u_a\Phi^a=u_{\widehat{0}}\Phi^{\widehat{0}}-\bm{\widehat{u}}\cdot\bm{\widehat{\Phi}}=0$, it satisfies the relation $\Phi^{\widehat{0}}=(\bm{\widehat{u}}\cdot\bm{\widehat{\Phi}})/u_{\widehat{0}}=\bm{\beta}\cdot\bm{\widehat{\Phi}}$, where $\bm\beta$ is defined by Eq. (\ref{momentum}). The similar relation for the four-spin is given by Eq. (\ref{spin}). We can perform the following transformation: %($\widehat{\bm\beta}=\bm{\widehat{u}}/u^{\widehat{0}}$):
\begin{equation}\begin{array}{c}
a_b \frac{du^b}{d\tau}=a_{\widehat{0}}\,\frac{du^{\widehat{0}}}{d\tau}-\widehat{\bm a}\cdot{\bm\beta}\,\frac{du^{\widehat{0}}}{d\tau}-u^{\widehat{0}}\,\widehat{\bm a}\cdot\frac{d{\bm\beta}}{d\tau}=-u^{\widehat{0}}\,\widehat{\bm a}\cdot\frac{d{\bm\beta}}{d\tau},~~~
u^a \frac{du^b}{d\tau}a_b=-u^a u^{\widehat{0}}\,\widehat{\bm a}\cdot\frac{d{\bm\beta}}{d\tau}.
\end{array} \label{Jackson2} \end{equation}

Thus, Eq. (\ref{BMT1}) leads to
\begin{equation}
\frac{da^{\widehat{0}}}{d\tau}=\Phi^{\widehat{0}}+\left(u^{\widehat{0}}\right)^2\,\widehat{\bm a}\cdot\frac{d{\bm\beta}}{d\tau}, ~~~ \frac{d\widehat{\bm a}}{d\tau}=\bm{\widehat{\Phi}}+\left(u^{\widehat{0}}\right)^2\,{\bm\beta}\left(\widehat{\bm a}\cdot\frac{d{\bm\beta}}{d\tau}\right).
\label{Jackson3} \end{equation}

Now we can calculate the equation of motion for the rest frame spin $\bm\zeta$ with the use of the relations
$$ %\begin{equation}
{\bm \zeta}=\widehat{\bm a}-\frac{u^{\widehat{0}}}{u^{\widehat{0}}+1}{\bm\beta}({\bm\beta}\cdot\widehat{\bm a}),~~~\frac{d}{d\tau}\left(\frac{u^{\widehat{0}}}{u^{\widehat{0}}+1}{\bm\beta}\right)=\frac{u^{\widehat{0}}}{u^{\widehat{0}}+1}\frac{d\bm\beta}{d\tau}+
\frac{\left(u^{\widehat{0}}\right)^3}{(u^{\widehat{0}}+1)^2}{\bm\beta}\left({\bm\beta}\cdot\frac{d\bm\beta}{d\tau}\right).
$$ %\label{reltn} \end{equation}
The needed equation has the form (cf. Refs. \cite{Jackson,PhysScr})
\begin{equation}
\frac{d\bm\zeta}{d\tau}=\bm{\widehat{\Phi}}-\frac{u^{\widehat{0}}\bm\beta}{u^{\widehat{0}}+1}\Phi^{\widehat{0}}+
\frac{\left(u^{\widehat{0}}\right)^2}{u^{\widehat{0}}+1}\bm\zeta\times\left({\bm\beta}\times\frac{d\bm\beta}{d\tau}\right).
\label{Jackson4} \end{equation}

The transformation of the given four-vector $\Phi^a$ to the instantaneously accompanying frame results in
$\bigl(\Phi^{(0)}\bigr)^a=\bigl(0,\bm{\widehat{\Phi}}^{(0)}\bigr)$, where
$$\bm{\widehat{\Phi}}^{(0)}=\bm{\widehat{\Phi}}-\frac{u^{\widehat{0}}}{u^{\widehat{0}}+1}\bm\beta(\bm\beta\cdot\bm{\widehat{\Phi}})
=\bm{\widehat{\Phi}}-\frac{u^{\widehat{0}}\bm\beta}{u^{\widehat{0}}+1}{\widehat{\Phi}}^0.$$

As follows from Eq. (\ref{timedil}), the derivation of $\bm{\widehat{\Phi}}^{(0)}$ from Eq. (\ref{Jacksong}) brings the equation of the spin motion to the form
\begin{equation}
\frac{d\bm\zeta}{d\widehat{t}}=-\frac{\bm{\mathcal{B}}^{(0)}}{u^{\widehat{0}}}\times\bm\zeta+
\frac{u^{\widehat{0}}}{u^{\widehat{0}}+1}\bm\zeta\times\left({\bm\beta}\times\frac{d\bm\beta}{d\tau}\right).
\label{Jackson5} \end{equation}
The angular velocity of the spin precession is given by
\begin{equation}
\widetilde{\bm{\Omega}}=-\frac{\bm{\mathcal{B}}^{(0)}}{u^{\widehat{0}}}-
\frac{u^{\widehat{0}}}{u^{\widehat{0}}+1}\left({\bm\beta}\times\frac{d\bm\beta}{d\tau}\right),~~~ \bm\Omega=\frac{u^{\widehat{0}}}{{u}^{0}}\widetilde{\bm{\Omega}}.
\label{angvelo} \end{equation}

Since $\widetilde{\bm{\Omega}}=\widetilde{\bm{\Omega}}_L+\widetilde{\bm{\Omega}}_T$, the angular velocity of the Thomas precession is equal to
\begin{equation}
\widetilde{\bm{\Omega}}_T=-
\frac{u^{\widehat{0}}}{u^{\widehat{0}}+1}\left({\bm\beta}\times\frac{d\bm\beta}{d\tau}\right)=-
\frac{1}{u^{\widehat{0}}(u^{\widehat{0}}+1)}\left(\widehat{\bm{u}}\times\frac{d\widehat{\bm{u}}}{d\tau}\right).
\label{multanv} \end{equation}
Explicitly,
\begin{equation}\begin{array}{c}
\widetilde{\bm{\Omega}}_T=-
\frac{\widehat{\bm{u}}\times\bm{\mathcal{E}}^{(0)}}{u^{\widehat{0}}(u^{\widehat{0}}+1)}=-
\frac{1}{u^{\widehat{0}}+1}\left[\widehat{\bm{u}}\times\bm{\mathcal{E}}+\frac{\widehat{\bm{u}}\times(\widehat{\bm{u}}\times\bm{\mathcal{B}})}
{u^{\widehat{0}}}\right]\\
=\frac{u^{\widehat{0}}-1}{u^{\widehat{0}}}\bm{\mathcal{B}}-\frac{1}{u^{\widehat{0}}(u^{\widehat{0}}+1)}\widehat{\bm u}(\widehat{\bm u}\cdot\bm{\mathcal{B}})-
\frac{\widehat{\bm u}\times\bm{\mathcal{E}}}
{u^{\widehat{0}}+1}.
\end{array} \label{Thpexpl} \end{equation}
Equation (\ref{omgemtt}) can be presented in terms of the rest frame fields:
\begin{equation}\begin{array}{c}
\widetilde{\bm{\Omega}}=-\frac{\bm{\mathcal{B}}^{(0)}}{u^{\widehat{0}}}-
\frac{\widehat{\bm{u}}\times\bm{\mathcal{E}}^{(0)}}{u^{\widehat{0}}(u^{\widehat{0}}+1)}.
\end{array} \label{omgfm}
\end{equation}
Equations (\ref{omgemtt}), (\ref{omgemOu}), (\ref{Thpexpl}), and (\ref{omgfm}) are consistent.

That, the use of the PK fields has allowed us to derive Eq. (\ref{multanv}) for the Thomas effect in Riemannian spacetimes. We have rigorously proven that this equation has practically the same form as the corresponding equation (\ref{Thompre}) defining the Thomas effect in electrodynamics. However, the equation for the Thomas effect in the world frame (with a substitution of the velocity and the acceleration in the world frame for the corresponding LLF quantities) may be different.

Let us calculate the contributions from the local Lorentz transformations and from the Thomas effect to the angular velocity of the spin precession for the uniformly accelerated frame and for the Schwarzschild field in the isotropic coordinates. We can use the weak-field approximation. For the uniformly accelerated frame, these contributions are given by
\begin{equation}\begin{array}{c}
\widetilde{\bm{\Omega}}_L^{(a)}=-\frac{c\widehat{\bm u}\times\bm a(t)}{c^2+\bm a(t)\cdot\bm r},~~~
\widetilde{\bm{\Omega}}_T^{(a)}=\frac{u^{\widehat{0}}}{u^{\widehat{0}}+1}\cdot\frac{c\widehat{\bm u}\times\bm a(t)}{c^2+\bm a(t)\cdot\bm r}.
\end{array} \label{omgaccf}
\end{equation}

In the weak-field approximation, this equation takes the form
\begin{equation}\begin{array}{c}
\widetilde{\bm{\Omega}}_L^{(a)}=-\frac{\widehat{\bm u}\times\bm a(t)}{c},~~~
\widetilde{\bm{\Omega}}_T^{(a)}=\frac{u^{\widehat{0}}}{c(u^{\widehat{0}}+1)}\widehat{\bm u}\times\bm a(t).
\end{array} \label{omgacwf}
\end{equation}

The corresponding relations for the Schwarzschild field in the isotropic coordinates are given by
\begin{equation}\begin{array}{c}
\widetilde{\bm{\Omega}}_L^{(i)}=\frac{2\widehat{\bm u}\times\bm g}{c},~~~
\widetilde{\bm{\Omega}}_T^{(i)}=-\frac{2\left(u^{\widehat{0}}\right)^2-1}{cu^{\widehat{0}}(u^{\widehat{0}}+1)}\widehat{\bm u}\times\bm g.
\end{array} \label{omgSchw}
\end{equation}
When $\bm a=const=-\bm g$, Eqs.  (\ref{omgacwf}) and (\ref{omgSchw}) significantly differ. We have discussed the origin of this difference in Sec. \ref{Comparison}. We should underline that neither the contributions from the local Lorentz transformations nor those from the Thomas effect vanish in the nonrelativistic limit. It has been claimed in Ref. \cite{Lee} that the spin rotation in the uniformly accelerated frame is caused only by the Thomas effect. The fallacy of this claim has been shown in Ref. \cite{MashhoonObukhov}.

We can also specify the two contributions to the angular velocity of the spin precession in the rotating frame. In this case, the both contributions are also nonzero and are given by
\begin{equation}\begin{array}{c}
\widetilde{\bm{\Omega}}_L=-\bm\omega u^{\widehat{0}}+\frac{\widehat{\bm u}(\widehat{\bm u}\cdot\bm\omega)}{u^{\widehat{0}}+1},~~~
\widetilde{\bm{\Omega}}_T=-\frac{\widehat{\bm{u}}\times(\widehat{\bm{u}}\times\bm{\omega})}{u^{\widehat{0}}+1}=\bm\omega (u^{\widehat{0}}-1)-\frac{\widehat{\bm u}(\widehat{\bm u}\cdot\bm\omega)}{u^{\widehat{0}}+1}.
\end{array} \label{omgrofm}
\end{equation}
The quantity
\begin{equation}\begin{array}{c}
\bm{O}=-\frac{1}{u^{0}(u^{0}+1)}\left(\bm{u}\times\frac{d\bm{u}}{d\tau}\right)
\end{array} \label{omgwm}
\end{equation}
frequently used for a specification of the Thomas precession in curved spacetimes can be obtained with Eq. (\ref{LorChrc}).
In the weak-field approximation ($\widehat{\bm u}\approx \bm u,~u^{\widehat{0}}=u^{0}$), the quantity $\bm{O}$ is twice as much as $\widetilde{\bm{\Omega}}_T$,
\begin{equation}\begin{array}{c}
\widetilde{\bm{\Omega}}_T=-\frac{\bm{u}\times(\bm{u}\times\bm{\omega})}{u^{0}+1},\end{array} \label{omgwa}
\end{equation}\begin{equation}\begin{array}{c} \bm{O}=-\frac{2\bm{u}\times(\bm{u}\times\bm{\omega})}{u^{0}+1}.
\end{array} \label{omgwn}
\end{equation}

This example unambiguously shows that the use of the quantity (\ref{omgwm}) for the determination of the Thomas precession in curved spacetimes is incorrect. We should nevertheless mention that Eqs. (\ref{multanv}),  (\ref{Thpexpl}),  (\ref{omgrofm}), and
 (\ref{omgwa}) describe the precession of the spin pseudovector defined in the LLF. These equations cannot be directly applied to the Thomas precession of a segment of a rapidly rotating disk which has been considered in Ref. \cite{Whitmire}.

%%%%%%%%%%%%%%%%%%%%%%%%%%%%%%%%%%%%%%%%%%%%%%%%%
\section{Discussion and summary}\label{final}
%%%%%%%%%%%%%%%%%%%%%%%%%%%%%%%%%%%%%%%%%%%%%%%%%

The introduction of the gravitoelectromagnetic fields being effective fields in an anholonomic tetrad frame (coframe) significantly simplifies a description of motion of spinning
%(as well as spinless)
particles in GR. When one neglects the spin-curvature coupling and the mutual influence of particle and spin motion, dynamics of the four-velocity and spin is defined by Eqs. (\ref{omgem}) and (\ref{force}) similar to corresponding equations in electrodynamics.  However, the equations of motion for the four-velocity and spin are not equally useful. The conventional three-component spin is defined in the particle rest frame which is one of LLFs. In contrast, the four-velocity is the world vector and its evolution should be defined in the world frame.
Certainly, the transition to the world coordinates is not difficult because
$$u^\mu=e^\mu_au^a,~~~\frac{du^\mu}{d\tau}=e^\mu_a\frac{du^a}{d\tau}+u^a\frac{de^\mu_a}{d\tau}.$$
Moreover, the investigation of the particle motion is simplified when the
trajectory is infinite. In this case, one can apply the fact that
the quantities $u^\mu$ and $u^a$ coincide at the initial and final parts of
the particle trajectory because of the very large distance to
the field source \cite{ostrong}. However, the use of Eq. (\ref{LorChrc}) seems to be more straightforward.

For a derivation of the equations of motion, canonical methods based on the use of Hamiltonians and Lagrangians can also be successfully applied.

Basic tetrads satisfy the Schwinger gauge while other tetrads are also applicable. Tetrads which do not satisfy this gauge are carried by observers moving in a described spacetime.
All tetrads are equivalent and the use of any tetrad is possible. Different tetrads are connected by local Lorentz transformations. This connection is determined in Sec. \ref{Lotrfra}. In the general case, the gravitoelectromagnetic fields differ in different coframes. In accordance with Refs. \cite{Arminjon,Arminjontalk}, these fields do not coincide even for different tetrads belonging to the Schwinger gauge (see the example given in Sec. \ref{curvilinearcoordinates}).

In the present work, we explain and develop the conception of gravitoelectromagnetism first proposed by Pomeransky and Khriplovich \cite{PK}. We calculate the gravitoelectromagnetic fields in the most important special cases (Sec. \ref{examples}) and determine their local Lorentz transformations (Sec. \ref{Lorentztransformations}). The validity of these transformations is demonstrated for the case of the uniformly accelerated frame.

We apply the gravitoelectromagnetic fields for a comparison of inertia and gravity and of the Mathisson and Stern-Gerlach forces. In agreement with Refs. \cite{VarjuRyder,Lee,PRD,MashhoonObukhov}, the uniformly accelerated frame cannot completely imitate the gravitational field of the Schwarzschild source. %The reason is the gravitomagnetic field which is zero in the uniformly accelerated frame but acts in the Schwarzschild field.
The forces conditioned by the mismatched terms in the equations of motion for the Schwarzschild field in the Cartesian and isotropic coordinates and for the uniformly accelerated frame are proportional to \emph{first} derivatives of the spatial components of the metric tensor while the tidal and Mathisson forces are defined by \emph{second} derivatives of the metric tensor. However, we cannot support the claim made in Refs. \cite{VarjuRyder,Lee} that these properties violate Einstein's equivalence principle. This principle has been formulated only relative to constant \emph{uniform} gravitational fields. The Schwarzschild field (as well as other real gravitational fields) is \emph{nonuniform}. We have shown that the spatial inhomogeneity significantly influences the form of the equations of motion. The terms different for the Schwarzschild field and the uniformly accelerated frame ($\bm g=-\bm a$) are of the same order of magnitude as the corresponding terms different for the Schwarzschild field in the Cartesian and isotropic coordinates.  Therefore, the difference between the equations of motion in the Schwarzschild field and the uniformly accelerated frame does not violate Einstein's equivalence principle.

An expression of the Mathisson force in terms of the gravitoelectromagnetic fields allows us to state that the deep similarity between the gravity and electromagnetism exists even for effects depending on curvature. It is well known that the Mathisson force violates the weak equivalence principle.

%The careful statement of the equivalence principle asserting the equivalence between inertia and gravity is always valid. This statement does not require an identity of inertial and %gravitational phenomena. The frequently used mathematical formulation of the equivalence principle is given by Eqs. (\ref{eqPKu}) and (\ref{eqPKn}). This formulation is valid when one %allows for first derivatives of the metric tensor but is violated by the Mathisson force.

Probably the most exciting result of the use of the local Lorentz transformations and the gravitoelectromagnetic fields is the general description of the Thomas precession in GR carried out in Sec. \ref{Thomasprecession}. Amazingly, Eq. (\ref{multanv}) defining the angular velocity of the Thomas precession in LLFs is analogous to the corresponding formula \cite{Thomas,Jackson,DraganThomas,Rindler,Stepanov,PhysScr} of special relativity.
Equations (\ref{Thpexpl}) and (\ref{omgfm}) show the convenience of the gravitoelectromagnetic fields for a description of spin effects in GR and detach the Thomas effect.

We underline a great importance of the Thomas effect for a better understanding of spin dynamics in inertial and gravitational fields. Experimental investigations of this dynamics in turn are very important to determine fundamental properties of gravity. In particular, the Gravity Probe B experiment \cite{GPB} has confirmed the theoretical prediction \cite{Schiff} (see also Sec. \ref{LenThir}) for the spin precession due to the geodetic and LT effects. The LT effect for orbiting bodies (frame dragging) has been certified in experiments with the LAGEOS satellites \cite{Ciufolini}. Experiments with atomic and nuclear spins in Earth's rotating frame \cite{Venema,Gemmel} have verified the behavioral equivalence of quantum mechanical spins and classical gyroscopes \cite{PRD2007,OSTORSION}. Experimental constraints for equivalence principles and new interactions have been analyzed in Ref. \cite{Ni}.

We can conclude that the conception of gravitoelectromagnetism used here perfectly describes the evolution of %four-velocity and
the spin of a relativistic particle in general noninertial frames and arbitrarily strong gravitational fields. The main distinctive features of this conception are comparatively simple equations of motion and the clear analogy between electromagnetism and gravity. The introduction of the gravitoelectromagnetic fields can be regarded as a comparatively
new powerful method in general relativity.

%%%%%%%%%%%%%%%%%%%%%%%%%%%%%%%%%%%%%%%%%%%%%%%%%
\section*{Acknowledgments}
%%%%%%%%%%%%%%%%%%%%%%%%%%%%%%%%%%%%%%%%%%%%%%%%%

The work was supported in part by the Belarusian Republican Foundation for Fundamental Research
(Grant No. $\Phi$16D-004) and by the Heisenberg-Landau program of the German
Ministry for Science and Technology (Bundesministerium f\"{u}r Bildung und
Forschung)).


\begin{thebibliography}{99}

\bibitem{MTW} C. W. Misner,
K. S. Thorne, and J. A. Wheeler, \emph{Gravitation}. Freeman, San
Francisco (1973).

\bibitem{rmp}
F. W. Hehl, P. von der Heyde, G. D. Kerlick, and J. Nester,
%{\it General relativity with spin and torsion: foundations and prospects},
Rev. Mod. Phys. {\bf 48} (1976) 393-416.

\bibitem{HehlTwoLectures}
F. W. Hehl, J. Lemke, and E. W. Mielke, Two lectures on fermions and gravity, in: \emph{Geometry and Theoretical Physics}, Proc. of the Bad Honnef School (12-16 Feb 1990), J. Debrus and A. C. Hirshfeld, eds. (Springer, New York, 1991), pp. 56-140.

\bibitem{Warszawa}
A. J. Silenko,
%Classical and quantum spins in curved spacetimes.
Acta Phys. Polon. B Proc. Suppl. \textbf{1}, 87 (2008). %-107.

\bibitem{OST}
Yu. N. Obukhov, A. J. Silenko, and O. V. Teryaev,
%Spin dynamics in gravitational fields of rotating bodies
%and the equivalence principle.
Phys. Rev. D \textbf{80}, 064044 (2009).

\bibitem{Schwinger}
J. Schwinger, %Energy and momentum density in field theory.
Phys. Rev.  \textbf{130}, 800 (1963); % 800-805;
%%%%%%%J. Schwinger, %Quantized gravitational field.
%%%%%%%Phys. Rev.
\textbf{130}, 1253 (1963). %1253-1258.

\bibitem{dirac}
P. A. M. Dirac, Interacting gravitational and spinor fields,
in: \emph{Recent developments in general relativity}, Festschrift for Infeld.
Pergamon Press, Oxford and PWN, Warsaw (1962), pp. 191-200.

\bibitem{Arminjon}
M. Arminjon, %On the Non-uniqueness Problem of the Covariant Dirac Theory and the Spin-Rotation Coupling
Int. J. Theor. Phys. % (2013) 52:4032-4044
\textbf{52}, 4032 (2013);
%Should There Be a Spin-Rotation Coupling for a Dirac Particle?
%Int. J. Theor. Phys. % (2014) 53:1993-2013
\textbf{53}, 1993 (2014);
%Some Remarks on Quantum Mechanics in a Curved Spacetime, Especially for a Dirac Particle
\textbf{54}, 2218 %-2235
(2014).

\bibitem{Arminjontalk}
M. Arminjon, %On the Hamiltonian and energy operators in a curved spacetime, especially for a Dirac particle
J. Phys.: Conf. Ser. \textbf{626}, 012030 (2015).

\bibitem{MashhoonGrEM}
B. Mashhoon, Gravitoelectromagnetism: A brief review, in:
\emph{The Measurement of Gravitomagnetism: A Challenging
Enterprise}. Edited by L. Iorio, Nova Science, New York (2007), pp.
29-39.

\bibitem{Tartaglia}
A. Tartaglia and M. L. Ruggiero, %Gravito-electromagnetism versus
%electromagnetism.
Eur. J. Phys. \textbf{25}, 203 (2004); %-210;
A. Tartaglia and M. L. Ruggiero, Analogies and Differences between Gravito-Electromagnetism and Electromagnetism, in:
\emph{The Measurement of Gravitomagnetism: A Challenging
Enterprise}. Edited by L. Iorio, Nova Science, New York (2007), pp. 41-50.

\bibitem{Ryder}
L. Ryder, \emph{Introduction to General Relativity}. Cambridge Univ. Press, New York (2009).

\bibitem{PK}
A. A.~Pomeransky and I. B.~Khriplovich,
%Equations of motion of spinning relativistic particle in external fields.
Zh.\ Eksp.\ Teor.\ Fiz.\ \textbf{113}, 1537 (1998) %-1557
[J.\ Exp.\ Theor.\ Phys.\ \textbf{86}, 839 (1998)]. %-849].

\bibitem{KO}
I. Yu. Kobzarev, L. B. Okun,
%Gravitational Interaction Of Fermions.
Zh. Eksp. Teor. Fiz. \textbf{43}, 1904 (1962) %-1909
[Sov. Phys. JETP \textbf{16}, 1343 (1963)]. %-1346].

\bibitem{T2} O. V.~Teryaev,
%Sources of time reversal odd spin asymmetries in QCD.
Czech.\ J.\ Phys.\ \textbf{53}, 47B (2003); %-58B;
arXiv:hep-ph/9904376. %arXiv:hep-ph/0306301.
%\bibitem{Ter2} Teryaev, O.V. Sources of time reversal odd spin asymmetries in QCD
%/ O.V. Teryaev // Czech. J. Phys. -- 2003. -- Vol. 53. -- Suppl.
%B. -- P. 47B-58B.

\bibitem{Mathisson}
M. Mathisson, %Neue Mechanik materieller Systeme.
Acta Phys. Polon. {\bf 6}, 163 (1937). %-200

\bibitem{Papapetrou}
A. Papapetrou,
%Spinning test-particles in general relativity. I.
Proc. Roy. Soc. Lond. A \textbf{209}, 248 (1951). %-258.
 %%CITATION = PRSLA,A209,248;%%

\bibitem{Plyatsko:1997gs}
R.~Plyatsko,
%Gravitational ultrarelativistic spin orbit interaction and the weak equivalence principle.
Phys.\ Rev.\ D \textbf{58}, 084031 (1998).

\bibitem{PlyatskoF}
R. Plyatsko and M. Fenyk, %Highly relativistic spinning particle in the
%Schwarzschild field: Circular and other orbits.
Phys. Rev. D \textbf{85}, 104023 (2012).

\bibitem{PlyatskoFn}
R. Plyatsko and M. Fenyk,
%Highly relativistic circular orbits of spinning particle in the Kerr field.
Phys. Rev. D \textbf{87}, 044019 (2013).

\bibitem{Mashhoon}
A. Gorbatsevich, %Zur Quantentheorie in nichtinertialen
%Bezugssystemen.
Experimentelle Technik der Physik
(Berlin) \textbf{27}, 529 (1979); %-535;
B. Mashhoon,  %Neutron interferometry in a rotating frame of reference.
Phys. Rev. Lett. \textbf{61}, 2639 (1988). %-2642.

\bibitem{ostrong}
Yu. N. Obukhov,  A. J. Silenko, and  O. V. Teryaev, %  Dirac
%fermions in strong gravitational fields.
Phys. Rev. D \textbf{84}, 024025 (2011).  %Iss. 2. P. .

\bibitem{OSTgrav} Y. N. Obukhov, A. J. Silenko, O. V. Teryaev, %Spin in an arbitrary gravitational field.
Phys. Rev. D \textbf{88}, 084014 (2013).

\bibitem{Thomas}
%Thomas L H 1926 The Motion of the Spinning Electron \emph{Nature (London)} \textbf{117} 514-514;
%Thomas L H 1927 The Kinematics of an Electron with an Axis \emph{Philos. Mag.} {\bf 3} 1-22
L. H. Thomas, Nature (London) \textbf{117}, 514 (1926);
Philos. Mag. {\bf 3}, 1 (1927).

\bibitem{Jackson}
J. D. Jackson, \emph{Classical Electrodynamics}. 3rd edn. John Wiley $\&$ Sons, New York  (1998).

\bibitem{DraganThomas}
%Dragan A and Odrzyg\'{o}\'{z}d\'{z} T 2013 A half-page derivation of the Thomas precession
%\emph{Am. J. Phys.} \textbf{81}, 631-632
A. Dragan A and T. Odrzyg\'{o}\'{z}d\'{z}, Am. J. Phys. \textbf{81}, 631 (2013).

\bibitem{Rindler}
W. Rindler, \emph{Relativity: Special, General, and Cosmological}. 2nd edn. Oxford Univ. Press, Oxford (2001), pp. 199-200;
R. U. Sexl and H. K. Urbantke, \emph{Relativity, Groups, Particles: Special Relativity and Relativistic Symmetry in Field and Particle Physics}.  Revised edn. Springer, New York (2001);
A. A. Ungar, \emph{Beyond the Einstein
Addition Law and its Gyroscopic Thomas Precession}. Kluwer Acad. Publ., New York (2002).

\bibitem{Stepanov}
S. S. Stepanov, % 2012 Thomas precession for spin and rod \emph{Phys. Part. Nuclei} {\bf 43}, 128-145
Phys. Part. Nuclei {\bf 43}, 128 (2012).

\bibitem{PhysScr}
A. J. Silenko, %Spin precession of a particle with an electric
%dipole moment: contributions from classical electrodynamics and from the Thomas effect.
Phys. Scr. \textbf{90}, 065303 (2015).

\bibitem{OSTORSION}
Yu.\,N. Obukhov,  A.\,J. Silenko, and  O.\,V. Teryaev,  %  Spin-torsion coupling and gravitational moments of Dirac fermions:
%Theory and experimental bounds //
Phys. Rev. D \textbf{90}, 124068 (2014); Int. J. Mod. Phys.: Conf. Ser.
\textbf{40}, 1660081 (2016).

\bibitem{Obzor}
A. A.~Pomeransky, R. A.~Senkov, and I. B.~Khrip\-lo\-vich,
%Spinning relativistic particles in external fields.
Usp. Fiz. Nauk \textbf{43}, 1129 (2000) %-1141
 [Phys. Usp. \textbf{43}, 1055 (2000)]. %-1066].

\bibitem{Dvornikov}
M. Dvornikov, %Neutrino spin oscillations in gravitational fields.
%International Journal of Modern Physics D
%Vol. 15, No. 7 (2006) 1017 1033
Int. J. Mod. Phys. D \textbf{15}, 1017 (2006).%-1033.
%hep-ph/0601095.

\bibitem{BaileyIsrael}
I. Bailey, W. Israel, %Lagrangian Dynamics of Spinning Particles and Polarized Media in General Relativity
Commun. Math. Phys. \textbf{42}, 65
%-82
(1975).

\bibitem{YeeBander}
K. Yee, M. Bander, %Equations of motion for spinning particles in external electromagnetic and gravitational fields
Phys. Rev. D \textbf{48},  2797 %-2799 NUMBER 6 15 SEPTEMBER 1993
(1993).

\bibitem{SteinhoffSchafer}
J. Steinhoff, S. Hergt, and G. Sch\"{a}fer, %Next-to-leading order gravitational spin(1)-spin(2) dynamics in Hamiltonian form
Phys. Rev. D \textbf{77}, 081501(R) (2008); J. Steinhoff, G. Sch\"{a}fer, and S. Hergt, %ADM canonical formalism for gravitating spinning objects
Phys. Rev. D \textbf{77}, 104018 (2008); J. Steinhoff, S. Hergt, and G. Sch\"{a}fer, %Spin-squared Hamiltonian of next-to-leading order gravitational interaction
Phys. Rev. D \textbf{78}, 101503(R) (2008);
J. Steinhoff, G. Sch\"{a}fer,
%Canonical formulation of self-gravitating spinning-object systems
Europhys. Lett. \textbf{87}, 50004 (2009).

\bibitem{BarausseRacineBuonanno}
E. Barausse, E. Racine, and A. Buonanno, %Hamiltonian of a spinning test particle in curved spacetime
Phys. Rew. D \textbf{80}, 104025 (2009).

\bibitem{Jantzen}
R.\,T. Jantzen, P. Carini and D. Bini, %The Many Faces of Gravitoelectromagnetism
Ann. Phys. (N.Y.)  \textbf{215}, 1 %-50
(1992).

\bibitem{Costa}
L.\,F.\,O. Costa, %TIDAL TENSOR APPROACH TO GRAVITOELECTROMAGNETISM
Int. J. Mod. Phys. A \textbf{24}, %Nos. 8 & 9 (2009) 1695-1699
1695 (2009).

\bibitem{Schafer}
G. Sch\"{a}fer, %Gravitomagnetic Effects
Gen. Rel. Grav. \textbf{36}, %No. 10, October 2004
2223 %-2235
(2014).
%%%\bibitem{PRD}
%%%A.~J. Silenko and O.~V. Teryaev, %{\it Semiclassical limit for Dirac
%particles interacting with a gravitational field},
%%%Phys. Rev. D {\bf 71}, 064016 (2005).

\bibitem{PRD}
A.~J. Silenko and O.~V. Teryaev, % Semiclassical limit for Dirac
% particles interacting with a gravitational field,
Phys. Rev. D {\bf 71}, 064016 (2005). %; arXiv:gr-qc/0407015.

\bibitem{PRD2007}
A. J. Silenko and O. V. Teryaev, % Equivalence principle and experimental tests of gravitational spin effects,
Phys. Rev. D {\bf 76}, 061101(R) (2007).

\bibitem{Pirani} F. Pirani, %On the Physical significance of the Riemann tensor.
Acta Phys. Polon. \textbf{15}, 389 (1956). %-405.

\bibitem{T} W. Tulczyjev, Acta Phys. Polon. \textbf{18}, 393 (1959).

\bibitem{Plyatsko2015}
R. Plyatsko, M. Fenyk, %Highly relativistic spin-gravity coupling for fermions
Phys. Rew. D \textbf{91}, 064033 (2015).

\bibitem{LL4}
V. B. Berestetskii, E. M. Lifshitz, and L. P. Pitaevskii, \emph{Quantum electrodynamics.} Course of Theoretical Physics. Vol. 4, 2nd edn. Butterworth-Heinemann, Oxford (1982).

\bibitem{hergt}
S. Hergt and G. Sch\"afer,
%Higher-order-in-spin interaction Hamiltonian for binary black holes
%from source terms of Kerr geometry in approximate ADM coordinates.
Phys. Rev. D \textbf{77}, 104001 (2008). % (15 pages).

\bibitem{HN}
F. W.~Hehl and W. T.~Ni,
%Inertial effects of a Dirac particle,
  Phys.\ Rev.\ D \textbf{42}, 2045 (1990). %-2048.
%%CITATION = PHRVA,D42,2045;%%

\bibitem{SurfInv}
A. J. Silenko, Poverkhnost', No. 3,  % pp. 65-73. Quantum-mechanical description of spin 1/2 particles and nuclei channeled in bent crystals
65 (2015) [J. Surf. Invest. \textbf{9}, 272 (2015)].
%2015, Volume 9, Issue 2, pp 272-279

\bibitem{LT}
H. Thirring, %Uber die Wirkung rotierender ferner Massen in der
%Einsteinschen Gravitationstheorie.
%On the Effect of Rotating Distant
%Masses in Einstein's Theory of Gravitation.
Phys. %ikalische
Z. %eitschrift
\textbf{19}  (1918), 33-39 [Gen. Rel. Grav. 16 (1984), 712-725]; %Correction to my paper: ``On the effect of rotating distant masses in Einstein's theory of gravitation''.
Phys. Z. \textbf{22}, 29 (1921) %-30
[Gen. Rel. Grav. \textbf{16}, 725 (1984)]; %-727];
J. Lense and H. Thirring, %Uber den Einfluss der Eigenrotation der
%Zentralkorper auf die Bewegung der Planeten und Monde nach der
%Einsteinschen Gravitationstheorie. [
%On the Influence of the Proper Rotation
%of Central Bodies on the Motions of Planets and Moons According to Einstein's
%Theory of Gravitation.
Phys. Z. \textbf{19}, 156 (1918) %-163
[Gen. Rel. Grav. \textbf{16}, 727 (1984)]. %-750]. %; English translation
%of both papers as well as their critical analysis is available in: B.
%Mashhoon, F.W. Hehl, and D.S. Theiss, {\it On the gravitational
%effects of rotating masses: The Thirring-Lense papers}, Gen. Relat.
%Grav. {\bf 16} (1984) 711-750.

\bibitem{Lee}
T.-Y. Lee, %Gravitational spin-orbit coupling and the equivalence principle
Phys. Lett. A \textbf{291}, 1 (2001). % 1-3

\bibitem{MashhoonObukhov}
B. Mashhoon and Yu. N. Obukhov, %Spin precession in inertial and gravitational fields
Phys. Rev. D \textbf{88}, 064037 (2013).

\bibitem{VarjuRyder}
K. Varj\'{u}, and L. H. Ryder, %The effect of Schwarzschild field on spin l/2 particles compared
%to the effect of a uniformly accelerating frame
Phys. Lett. A \textbf{250}, 263 (1998).

\bibitem{DH}
J.~H. Donoghue and B.~R. Holstein,
%``Quantum mechanics in curved space,''
Am. J. Phys.  {\bf 54}, 827 (1986).

\bibitem{Schiff}
L. I. Schiff, %On Experimental Tests of the General Theory of
%Relativity,
%Am. J. Phys. \textbf{28}, 340 %-343,
%(1960); Proc. Nat. Acad. Sci. \textbf{46}, 871 (1960);
Phys. Rev. Lett. %Possible New Experimental Test of General Relativity Theory
\textbf{4}, 215 (1960).

\bibitem{Nicolaevici}
N. Nicolaevici, %Semi-Classical Derivation of the Spin-Orbit Coupling
%for the Dirac Particle in an Accelerated Frame
Gen. Rel. Gravit. \textbf{35}, %No. 11, November 2003
2017 %-2023
(2003).

\bibitem{Einstein}
A. Einstein, % \"{U}ber das Relativit\"{a}tsprinzip und die aus demselben gezogenen Folgerungen
Jahrbuch der Radioaktivitat \textbf{4}, 411 (1907); %-462
%  Über den Einfluss der Schwerkraft auf die Ausbreitung des Lichtes
Annalen Phys., \textbf{35}, 898 %-908.
(1911);
% Lichtgeschwindigkeit und Statik des Gravitationsfeldes,
Annalen Phys., \textbf{38}, 355 %-369.
(1912).

\bibitem{LL2}
L.D. Landau and E.M. Lifshitz, \emph{The Classical Theory of Fields},
4th edn (Butterworth-Heinemann, Oxford, 1980), p. 353.

\bibitem{Cogn}
G. Cognola, L. Vanzo, and S. Zerbini,
%Relativistic wave mechanics of spinless particles in a curved space-time.
Gen.\ Rel.\ Grav.\ \textbf{18}, 971 (1986). %-982.

\bibitem{Whitmire}
D. P. Whitmire, %Thomas Precession and the Relativistic Disk,
Nature Phys. Sci. \textbf{235}, 175 (1972).

%\bibitem{Mashhoonbook}
%\textcolor{blue} {B. Mashhoon, The Hypothesis of Locality and its Limitations, in: \emph{Relativity in Rotating Frames:
%Relativistic Physics in Rotating Reference Frames}, G. Rizzi, M. L. Ruggiero, eds. (Kluwer Acad. Publ., Dordrec (Kluwer Acad. Publ., Dordrecht, 2004), %pp. 43-55.}

\bibitem{GPB}
%Gravity Probe B: Final Results of a Space Experiment to Test General Relativity
C. W. F. Everitt, D. B. DeBra, B. W. Parkinson, J. P. Turneaure, J. W. Conklin, M. I. Heifetz, G. M. Keiser,
A. S. Silbergleit, T. Holmes, J. Kolodziejczak, M. Al-Meshari, J. C. Mester, B. Muhlfelder, V. G. Solomonik,
K. Stahl, P. W. Worden, Jr., W. Bencze, S. Buchman, B. Clarke, A. Al-Jadaan, H. Al-Jibreen, J. Li, J. A. Lipa,
J. M. Lockhart, B. Al-Suwaidan, M. Taber, and S. Wang, Phys. Rev. Lett. \textbf{106}, 221101 (2011).

\bibitem{Ciufolini}
I. Ciufolini and E. C. Pavlis, Nature \textbf{431}, 958 (2004).

\bibitem{Venema}
B.~J. Venema, P.~K. Majumder, S.~K. Lamoreaux, B.~R. Heckel, and E.~N. Fortson,
% Search for a coupling of the Earth's gravitational field to
%nuclear spins in atomic mercury,
Phys. Rev. Lett. {\bf 68}, 135 (1992).

\bibitem{Gemmel}
C. Gemmel, W. Heil, S. Karpuk, K. Lenz, Ch. Ludwig, Yu. Sobolev, K. Tullney, M. Burghoff, W. Kilian,
S. Knappe-Gr\"{u}neberg, W. M\"{u}ller, A. Schnabel, F. Seifert, L. Trahms, and St. Baessler,
%{\it Ultra-sensitive magnetometry based on free precession of nuclear spins},
Eur. Phys. J. D {\bf 57}, 303 (2010); %-320
%
%\bibitem{Gemmel2}
C. Gemmel, W. Heil, S. Karpuk, K. Lenz, Yu. Sobolev, K. Tullney, M. Burghoff, W. Kilian,
S. Knappe-Gr\"{u}neberg, W. M\"{u}ller, A. Schnabel, F. Seifert, L. Trahms, and U. Schmidt, %{\it Limit on Lorentz and CPT violation of the bound neutron
%using a free precession ${}^3$He/${}^{129}$Xe comagnetometer},
Phys. Rev. D {\bf 82}, 111901(R) (2010);
%
%\bibitem{burghoff}
M. Burghoff, C. Gemmel, W. Heil, S. Karpuk, W. Kilian, S. Knappe-Gr\"{u}neberg, K. Lenz, W. M\"{u}ller, K. Tullney, U. Schmidt, A. Schnabel, F. Seifert, Yu. Sobolev and L. Trahms, %{\it Probing Lorentz invariance and other fundamental
%symmetries in ${}^3$He/${}^{129}$Xe clock-comparison experiments},
J. Phys.: Conf. Ser. {\bf 295}, 012017 (2011);
W. Heil, C. Gemmel, S. Karpuk, Yu. Sobolev, K. Tullney, F. Allmendinger, U. Schmidt, M. Burghoff, W. Kilian, S. Knappe-Gr\"{u}neberg,
A. Schnabel, F. Seifert, and L. Trahms, Ann. Phys. (Berlin) \textbf{525}, %No. 8–9,
539 %–549
(2013); K. Tullney, W. Heil, S. Karpuk, Yu. Sobolev, F. Allmendinger, and U. Schmidt, Int. J. Mod. Phys. Conf. Ser. \textbf{40}, 1660083 (2016). %[6 pages] DOI: http://dx.doi.org/10.1142/S2010194516600831
%Search for Spin-Dependent Short-Range Interaction with an 3He/129129Xe Clock Comparison Experiment

\bibitem{Ni}
W. T.~Ni,
%Equivalence Principles and Electromagnetism,
Phys. Rev. Lett. \textbf{38}, 301 (1977);
%Searches for the role of spin and polarization in gravity,
Rep. Prog. Phys. \textbf{73}, 056901 (2010); % (24pp)
%Rotation, the Equivalence Principle, and the Gravity Probe B Experiment,
Phys. Rev. Lett. \textbf{107}, 051103 (2011);
%Searches for the Role of Spin and Polarization in Gravity: A Five-Year Update,
Int. J. Mod. Phys.: Conf. Ser. \textbf{40}, 1660010 (2016).

\end{thebibliography}
\end{document}